\documentclass[submit]{aiaa-tc} 
\usepackage{calligra}
\usepackage{color}
\usepackage{soul}

\usepackage{bm}
\usepackage{amsmath}
\usepackage{subfigure}
\usepackage[colorlinks=true, pdfstartview=FitV, linkcolor=black, citecolor= black, urlcolor= black]{hyperref}
\usepackage{overcite}
\usepackage{footnpag}                   
\usepackage{psfrag}
\usepackage{algorithm}\usepackage{mathtools}
\usepackage{algorithmic}
\usepackage{amssymb}

\usepackage{amsmath}
\usepackage{graphicx}

\title{Deep Reinforcement Learning for Six Degree-of-Freedom Planetary Powered Descent and Landing}

\author{%
Brian Gaudet\thanks{Research Engineer, Department of Systems and Industrial Engineering, Department of Aerospace and Mechanical Engineering}
\\
{\normalsize\itshape
   University of Arizona, Tucson Arizona, 85721, USA}\\
  Richard Linares%
   \thanks{Charles Stark Draper Assistant Professor, Department of Aeronautics and Astronautics, Email: linaresr@mit.edu.}    
    \\
  {\normalsize\itshape
   Massachusetts Institute of Technology, Cambridge, MA, 02139, USA}\\
   Roberto Furfaro\thanks{Professor, Department of Systems and Industrial Engineering, Department of Aerospace and Mechanical Engineering, E-mail: robertof@email.arizona.edu}
       \\
  {\normalsize\itshape
   University of Arizona, Tucson Arizona, 85721, USA}\\
}

\begin{document}
\maketitle{}

\begin{abstract}
Future Mars missions will require advanced guidance, navigation, and control algorithms for the powered descent phase to target specific surface locations and achieve pinpoint accuracy (landing error ellipse $<$ 5 m radius). The latter requires both a navigation system capable of estimating the lander's state in real-time and a guidance and control system that can map the estimated lander state to a commanded thrust for each lander engine. In this paper, we present a novel integrated guidance and control algorithm designed by applying the principles of reinforcement learning theory. The latter is used to learn a policy mapping the lander's estimated state directly to a commanded thrust for each engine, with the policy resulting in accurate and fuel-efficient trajectories. Specifically, we use proximal policy optimization, a policy gradient method, to learn the policy. Another contribution of this paper is the use of different discount rates for terminal and shaping rewards, which significantly enhances optimization performance. We present simulation results demonstrating the guidance and control system's performance in a 6-DOF simulation environment and demonstrate robustness to noise and system parameter uncertainty.
\end{abstract}

\section{Introduction}
Future unconstrained, science-driven, robotic and human missions to large and small planetary bodies will require a high degree of landing accuracy. Indeed, the next generation of planetary landers will need more advanced guidance and control capabilities to satisfy the increasingly stringent accuracy requirements. The latter is driven by the desire to explore regions on planets (e.g. Mars) and satellites (e.g. Moon) that have the potential to yield the highest scientific return. In the case of Mars, the most demanding mission segment is the probably the powered descent phase, where the goal is to achieve a soft pinpoint landing, which we will define as the norm of the position error less than 5 m and the magnitude of the landing velocity below 2 m/s, with the velocity vector directed primarily downward, and negligible deviation from an upright attitude and zero rotational velocity. During a typical Entry, Descent and Landing (EDL) as implemented in past robotic missions to Mars [\citenum{shotwell2005phoenix,braun2007mars}], the lander's sensors (radar altimeters) are effectively blind until the heat shield is jettisoned, at which point the lander's guidance, navigation, and control system must quickly estimate the lander's state from a prior distribution extending several km downrange and crossrange, and then use the sequence of state estimates to achieve a soft landing at the target position, typically within one minute of the heat shield being jettisoned. 

There are two necessary capabilities to enable a lander to achieve a pinpoint landing.  First, the navigation system must be capable of accurately estimating the lander's state during the powered descent phase.  Second, the lander's guidance and control system must be capable of using these state estimates to achieve a pinpoint landing. In previous work [\citenum{gaudet2014navigation}], we made  progress towards the development of a guidance, navigation, and control system enabling pinpoint landing for the Mars powered descent phase by developing a navigation system using a Rao-Blackwellized particle filter, a digital terrain map, and the radar altimeters that are currently used on Mars landers, and demonstrated pinpoint landing accuracy in 3-DOF Monte Carlo simulations by coupling the navigation system with the energy-optimal closed-loop guidance algorithm as developed by Battin [\citenum{battin1999introduction}, page 558] and D'Souza [\citenum{d1997optimal}] (here referred to as DR/DV algorithm). However, as the simulations were run in a 3-DOF environment, this work neglected one of the more challenging portions of the powered descent problem, mapping the guidance system's commanded acceleration to actuator commands. In
this work we focus on developing an integrated guidance and control system suitable for the powered descent phase.

For the case of the robotic exploration of Mars, landing accuracy is important for several reasons. First, given accurate high-resolution maps of the Mars surface, the pinpoint-landing problem subsumes the hazard avoidance problem, as a hazard-free landing site can be targeted. Second, delivering a rover closer to a location of scientific interest reduces the risk of the rover malfunctioning before it reaches the desired site. In fact, some sites might be inaccessible to a rover unless the lander can deliver the rover with pinpoint accuracy. Moreover, reducing the distance the rover is required to travel can relax the design requirements for the rover, with a potential reduction in rover mass. Clearly, landing accuracy on the order of several meters is desirable as it would both reduce mission risk and extend the scope of feasible missions.

Considering guidance and control systems that have been deployed for an actual Mars landing, the current benchmark for landing accuracy is the Mars Science Laboratory (MSL), which used a velocity trigger for chute deployment and had an expected 3-sigma landing error ellipse of 19 km by 7 km [\citenum{steltzner2010mars}]. This landing ellipse is significantly reduced as compared to previous missions (e.g. [\citenum{shotwell2005phoenix}]), largely due to active bank angle control during the hypersonic re-entry phase based off of the lander's position as estimated by the lander's inertial measurement unit. Using a range trigger for chute deployment has the potential to improve parachute deployment accuracy, which could potentially give landing accuracy on the order of 4-6 km [\citenum{acikmese2012g}]. While this is a significant improvement, it is still a long way from pinpoint accuracy. The MSL system used a trajectory commander to generate a multi-phase trajectory at the start of the powered descent phase. This trajectory is designed so that it can be tracked by the trajectory follower system, which employs six feedback control loops, taking position and attitude estimates from the navigation filter.  Note that the trajectory was not optimized for minimal fuel usage [\citenum{singh2007guidance}]. One reason for the poor landing accuracy is that the lander is only capable of estimating altitude and velocity during the powered descent phase using the lander's radar altimeters. Moreover, even if the lander could accurately estimate its position and velocity in the target-centered reference frame in real time, the guidance law is only capable of relatively small divert maneuvers [\citenum{acikmese2012g}].

More recently, a method of loss-less convexification of non-convex control problems has been applied to the planetary landing problem [\citenum{accikmecse2013lossless},\citenum{acikmese2007convex}].  This work proposes a method to convexify the constraints of minimum thrust and thrust vector direction in such a way that the optimal solution of the convexified problem is optimal for the original problem. Convexification allows the use of interior point solution methods that guarantee a solution in bounded polynomial time, making it suitable for real-time powered descent. A cone constraint is imposed to enforce glide slope constraints and avoid terrain features. The trajectory is optimized to minimize fuel usage, and can target a desired landing site from any state in the deployment ellipse, thereby allowing pinpoint landing if coupled with a suitable navigation system. Importantly, the optimization problem is defined with three degrees of freedom, with the lander modeled as a point mass. Consequently, the trajectory would need to be mapped to lander engine thrust commands using a separately optimized controller. Although this is a significant improvement over prior methods, this approach may not result in the most fuel-efficient trajectories, as the optimization does not take into account atmospheric drag and the aerodynamics of the lander. Moreover, the approach may not be suitable for missions where a cone constraint on the lander's position is incompatible with mission goals.  One example of such a mission is landing at the bottom of a deep crater, where a cone  cannot be defined that  encloses the lander's position at the start of the powered descent phase, has an apex a the desired landing site, and avoids the crater walls. Another limitation is that future missions might require constraints where loss-less convexification is not possible. 

In  Ref. \citenum{szmuk2018successive}, the loss-less convexification approach was extended to generate 6-DOF trajectories through successive convexifications.  Starting with an initial guess for the trajectory, the dynamics are linearized around the trajectory using trust-regions to ensure that the problem remains bounded and feasible at each successive convexification. The successive convexifications continue until convergence, which is defined using tolerance thresholds. To make the problem tractable, the authors assume a constant inertia tensor, although in general, the inertia tensor will change as fuel is consumed.  Importantly, this approach will require either a separate controller to track the open loop 6-DOF trajectory or the trajectory will have to be recomputed at each time step as in model predictive control. The paper presents simulation results using a lander with a single gimbaled thruster in an environment which may not be representative of a specific mission.

In Ref. \citenum{lu2017propellant} the author uses an optimal control formulation to solve a 3-DOF Mars powered descent problem. The method, named Universal Powered Guidance (UPG), finds a closed form solution for the thrust amplitude and number of switches for the thrust profile and uses a numerically robust zero-finding algorithm to determine the initial costates as well as the optimal burn times. When compared to the Guidance for Fuel-Optimal Large Divert (G-FOLD, [\citenum{dueri2016customized,blackmore2010minimum,accikmecse2013lossless}]) algorithm  which is based on the above-mentioned convexification methodology,the UPG code is generally simpler and requires no special customization of the optimization algorithm. Additionally, it is capable of accommodating a variety of problem formulations which may not be possible via a second-order coning approach. The approach has the additional advantage of computing an optimal state at which to start the powered descent phase such that divert requirements are minimized. A disadvantage of this approach is that  constraints on glideslope and thrust vector direction are difficult to enforce. Recently, UPG has been recently employed to evaluate the interplay occurring between entry trajectory and fuel usage during a human mission to Mars [\citenum{lu2018adaptive}].

To summarize, current practice and the majority of proposed powered decent phase guidance and control systems use two separate and independently optimized systems for guidance and control. The navigation system estimates the lander's state from sensor measurements and passes this estimate to a guidance system.  This guidance system generates a trajectory that in general specifies the lander's target state as a function of time.  The trajectory is then passed to a control system that tracks the trajectory by determining which thrusters to fire and at what thrust level. There are variations on this approach that, to date, are primarily used for missile guidance.  For example, instead of generating a trajectory in real time based off of the lander's state at the beginning of the powered descent phase, a global guidance  law such as DR/DV [\citenum{d1997optimal}] can be developed that maps the lander's state in the target-centered reference frame to a commanded acceleration in the inertial frame. This commanded acceleration is then passed to an actuator control system (sometimes referred to as the autopilot), which must map the commanded acceleration in the inertial frame to a command to fire one or more of the lander's thrusters. Regardless of whether the approach uses trajectory generation coupled with a tracking controller or a guidance law coupled with an autopilot, the basic structure of three separately optimized systems (guidance, navigation, and control) is currently the state-of-the-art.

There are multiple issues that can arise with the separate optimization of the guidance and control systems. First, traditional autopilot design is typically simplified by having three separate controllers, one for roll stabilization, and the other two adjusting the actuators to modify yaw and pitch to attempt to match the commanded acceleration [\citenum{yanushevsky2007modern}]. When an autopilot is split into multiple controllers, it is possible that the combined commanded thrust results in a thrust vector with components that partially cancel, thus reducing fuel efficiency. Autopilot controllers that decouple different types of motion can also limit the types of maneuvers a vehicle can make, which in turn slows the flight control response time. Regardless of the autopilot architecture, a common problem is actuator saturation, where the guidance system's optimal commanded acceleration cannot be implemented given actuator limitations. Although a well-designed control system can compensate for actuator saturation, performance is typically degraded. A complication of separate guidance and control system design is that the reference trajectory must ensure that the sequence of acceleration commands in $\mathbb{R}^{3}$ is implementable given the thruster configuration of the lander. As an extreme example, if the commanded thrust were to shift by a large angle in a short amount of time, the lander may not be able to change its attitude fast enough to track the commanded thrust. Separate optimization of guidance and control systems can also complicate the design process, as performance issues may only be revealed when the two systems are combined in simulation, requiring a redesign of one or more of the separate systems.

Integrated Guidance and Control (IGC) is an alternative to separate guidance and control systems that have the potential to improve system performance [\citenum{xin2006integrated}]. With IGC, the guidance and control systems are treated as a single system for design purposes. Recent work on IGC has focused primarily on the problem of missile homing phase guidance for both endoatmospheric and exoatmospheric intercepts [\citenum{yan2012integrated},\citenum{palumbo2004integrated},\citenum{xin2006integrated},\citenum{shtessel2003integrated}]. Recently, an IGC system for Mars landing using a disturbance observer and multiple sliding surface techniques has been proposed [\citenum{zhang2017integrated}], although only a single trajectory was simulated to verify the approach. Therefore it is not clear if the chosen parameters in the gain matrix would generalize to different initial conditions in the deployment ellipse. Moreover, the dynamics used to derive the guidance law do not take into account the lander's changing mass during the powered descent phase.  Finally, the system is not truly integrated as the authors assume the ability to directly set the torque and the ability to generate translational thrust in arbitrary directions in $\mathbb{R}^{3}$ (not possible for a realistic thruster configuration).  This essentially assumes decoupling of translational and rotational motion; consequently, the system does not map observations directly to commanded thrust for the lander's individual engines, and a separate controller would be required to do so.  Wibben and Furfaro use a multiple sliding surface approach to develop an IGC system for lunar landing [\citenum{wibben2012integrated}].  However, the system was not truly integrated, as the 3-DOF thrust vector is first computed and mapped to the body frame, where the algorithm then works out the required attitude and the required torque to target that attitude. Consequently, the system requires a separate controller to map commanded thrust in $\mathbb{R}^{3}$ and commanded torque to individual engine thrust commands. Moreover, the method was only demonstrated for a single trajectory (there were no Monte Carlo simulations), and the gain matrix had to be tuned to ensure the lander touched down in an upright position.  Consequently, it is not clear if that gain matrix would work over a full deployment ellipse.

In this work, we develop an integrated guidance and control system that learns a global policy over the region of state space defined by the deployment ellipse and potential landing sites.  This global policy  maps the navigation system's estimate of the lander's state directly to commands specifying thrust levels for each engine. The global policy is learned using reinforcement learning (RL). Learning involves simulated interaction between an agent instantiating the policy and the environment over many episodes with randomly generated initial conditions that cover the deployment ellipse. Constraints such as minimum and maximum thrust, glide slope, maximum attitude and rotational velocity, and terrain feature avoidance (such as targeting the bottom of a deep crater) can be included in the reward function, and will be accounted for when the policy is optimized. Note that there is no guarantee on the optimality of trajectories induced by the policy, although in practice it is possible to get close to optimal performance by tuning the reward function. 

RL algorithms can be broken down into two major classes, i.e., value function methods and policy gradient methods.  Value function-based algorithms learn a mapping between a state - action tuple and the sum of the future discounted rewards received when starting in that specific state and taking that specific action. Actions with the highest value are then greedily chosen, with limited exploration achieved by choosing a random action with some small probability. Since all possible actions must be considered, these methods only work with discrete action spaces.  First proposed by Watkins [\citenum{watkins1992q}], recent work in value function-based algorithms include deep Q networks [\citenum{mnih2015human}]. The latter use experience replay to remove temporal correlation between samples and target networks which results in a reduced instability of combining bootstrapping with non-linear function approximators.  Deep Q networks have proven effective at control tasks requiring the mapping of pixel level observations directly to control actions, as demonstrated in [\citenum{mnih2015human}], although they are less effective at control problems where the dimensionality of the dynamics is larger, such as problems in robotic control, and are not suitable for problems requiring continuous action spaces.  Deterministic policy gradient algorithms, first proposed by Silver et al., [\citenum{silver2014deterministic}], and including the DDPG algorithm [\citenum{lillicrap2015continuous}], can be employed as an alternative to methods based on Deep Q learning. These algorithms also learn a mapping between state-action tuples and values.  However, rather than taking actions that globally maximize the value function, the algorithm maintains a separately parameterized policy mapping states to actions, and uses the chain rule to follow the gradient of the value function with respect to the action, given that the action is a function of the state as parameterized by the policy, 

Generally, policy gradient algorithms learn a direct stochastic mapping between an observation and action, with the policy network's parameters updated such that the probability of actions leading to higher future rewards is increased. First proposed by R. Williams [\citenum{williams1992simple}], these algorithms suffer from high variance, which can be substantially reduced by using a state value function baseline. With a baseline, actions that result in future discounted rewards higher than the estimated value of being in that state, as given by the state value function, have their probability increased. In this approach, a state value function mapping a state to the expected sum of future discounted rewards is learned concurrently with a stochastic policy mapping states to actions. For continuous action spaces, it is common to parameterize the policy as a Gaussian distribution with a diagonal covariance matrix, where the policy outputs the mean and variance of the actions conditioned on the state.  Recently, Shulman et. al. [\citenum{schulman2015trust}] have proven that policy gradient methods with stochastic policies can have monotonic improvement guarantees, provided that the policy changes during optimization as measured by the KL divergence are constrained to be within certain bounds. The policy optimization problem is posed as a constrained optimization problem that ensures the KL divergence between policy updates remains within specified bounds. The authors used this result to develop the trust region policy optimization algorithm (TRPO), which has proven effective at solving high-dimensional control problems. Later, Schulman [\citenum{schulman2017proximal}] proposed the proximal policy optimization (PPO) method that uses a heuristic to keep KL divergence between policy updates at a level that in practice ensures monotonic improvement during optimization. In practice, PPO works slightly better than TRPO, and has the advantage of not requiring second-order optimization methods. Policy gradient methods are much less sample efficient than methods based on Deep Q learning as they operate on policy. The latter means that policy gradient methods cannot take advantage of sample reuse through techniques such as experience replay [\citenum{mnih2015human}]. However, policy gradient methods are more stable, and work with minimal hyperparameter tuning, and tend to perform better in systems with high dimensional dynamics. 

 

A potential drawback of the RL approach is the lack of local or global stability guarantees.  In our opinion, this is not a major problem, as competing approaches such as optimal control, traditional linear control, and Lyapunov control only have stability guarantees if the system model is an accurate representation of the actual dynamic system.  In some cases, this assumption may hold, but in others, such as hypersonic re-entry, it may not be possible to apply these methods using a high fidelity system model. Nevertheless, the optimized guidance and control system could be tested using a high fidelity model. Even for the relatively simple dynamics of the Mars powered descent phase, recent works assume either a static inertia tensor [\citenum{szmuk2018successive}], or a static mass [\citenum{zhang2017integrated}]. In contrast, RL allows both optimization and verification using a high fidelity dynamics model. Since RL can be model free (as in this work), the complexities of the dynamics do not complicate the problem formulation within the RL framework. Another issue with RL is implementing hard state constraints. It is possible to approximate hard state constraints by terminating the episode and giving the agent a large negative reward upon a constraint violation, but the latter typically requires  manual tuning of the reward function, as we discuss in the sequel (see section II.D). Such manual tuning typically requires a certain amount of trial and error and results in the constraints being satisfied. However, there has been recent work on implementing hard constraints in RL using constrained policy optimization [\citenum{achiam2017constrained}]. Note that hard constraints on actions are simple; we clip any action returned by the policy such that it satisfies the constraints, and the agent learns to deal with this clipping. 

A comparison of RL and optimal control approaches to guidance and control are given below in Table~\ref{tab:rl_vs_oc}. The point of the comparison is not to make the case that one approach should be preferred over the other, but rather to suggest the scenarios where it might make sense to use the RL framework to solve guidance and control problems. 

\begin{table}[!ht]
	\fontsize{10}{10}\selectfont
    \caption{A Comparison of Optimal Control and RL.}
   \label{tab:rl_vs_oc}
        \centering 
   \renewcommand{\arraystretch}{1.5}
   \begin{tabular}{p{7cm} | p{7cm} } 
      \bf{Optimal Control} & \bf{Reinforcement Learning}\\
      \hline
      Single trajectory (except for trivial cases where HJB equations can be solved) & Global over theatre of operations \\
      Unbounded run time except for special cases such as convex constraints & Extremely fast run time for trained policy ($<$ 1ms in this work) \\
      Hybrid dynamics requires special treatment & Hybrid Dynamics handled seamlessly \\
      Dynamics need to be represented as ODE, possibly constraining fidelity of model used in optimization & No constraints on dynamics representation.  Agent can learn in a high fidelity simulator (i.e., Navier-Stokes modeling of aerodynamics) \\
      Open Loop (requires a controller to track the optimal trajectory) & Closed Loop (Integrated guidance and control) \\
      Output feedback (co-optimization of state estimation and guidance law) an open problem for non-linear systems & Can learn from raw sensor outputs allowing fully integrated GNC (pixels to actuator commands). Can learn to compensate for sensor distortion.\\
      Requires full state feedback &   Does not require full state feedback \\
      Elegantly handles state constraints & State constraints handled either via large negative rewards and episode termination or more recently, modification of policy gradient algorithm. Control constraints straightforward to implement \\
      Deterministic & Stochastic, learning does not converge every time, may need to run multiple policy optimizations\\
   \end{tabular}
\end{table}

Previous work using RL to solve control problems has focused on applications in robotics, with very few published works addressing problems in aerospace guidance and control. The first application of RL to a problem in the aerospace domain was applied to autonomous helicopter flight [\citenum{ng2003shaping}].  More recently, in Ref. \citenum{waslander2005multi} the authors compared reinforcement learning to sliding mode control and linear control in a quadrotor control application. The authors used policy iteration [\citenum{sutton1998reinforcement}] in a model-based formulation, where a model was learning from flight data using weighted least squares. Both the sliding mode and RL methods resulted in stable controllers, whereas the linear control methods failed.  

To our knowledge, this is the first published work demonstrating an RL-derived integrated guidance and control system applied to planetary landing in a 6-DOF environment. In fact, we were unable to find any work using RL to learn even a 3-DOF guidance law. In previous work, we achieved good results in a 2-DOF environment, but over a limited range of initial conditions [\citenum{gaudet2014adaptive}]. Open-AI gym has one environment for a 3-DOF planar lunar landing\footnote{https://gym.openai.com/envs/LunarLander-v2/}, but the problem is simplified by using a range of initial conditions that are unrealistically reduced from that required for an actual lunar landing (the crossrange dispersion is only three times the width of the landing site). In our work, we have found that solutions that work over a limited range of initial conditions often fail when the initial condition range is extended. Our ultimate goal is to use RL to develop a fully integrated guidance, navigation, and control system, where the policy maps raw sensor observations directly to actuator commands, i.e., radar altimeter readings to commanded thrust for the lander's individual engines. The RL framework can be applied to solve many different types of guidance and control problems, including missile homing phase guidance, exoatmospheric intercepts, hypersonic reentry maneuvering for planetary landings, and booster recovery via powered landing.

This paper is organized as follows. In section II, the problem formulation together with the RL methodology is applied to the proposed work. In section III, we show the results from policy optimization and testing for 6-DOF planetary landing scenarios. In section IV, a comparison between RL-based 3-DOF/6-DOF closed-loop policy and open-loop optimal trajectory derived using direct methods is reported. In section V, conclusions and future work are reported.

\section{Problem Formulation}

\subsection{Problem Statement}

Consider the powered descent problem on a large planetary body such as Mars. In this section we describe the RL set-up that is employed to find a 6-DOF closed-loop policy that generates quasi-fuel optimal trajectories capable of a) driving the lander to the desired location to the planetary surface with pin-point accuracy and b) satisfying flight and systems constraints (e.g. glide slope, thrust and attitude constraints). To set the stage for the IGC RL , consider the standard formulation for a typical trajectory optimization problem (see for example [\citenum{acikmese2007convex}]):

\textit{Minimum-Fuel Problem}: Find the thrust program and flight time that minimize the following cost functions:
\begin{equation}
\underset{t_f,\bf{T}}{\text{min}} \quad m_L(t_f)=\underset{t_f,{\bf T}}{\text{max}}\int_{0}^{t_f}\|\bf{T}\|dt
\end{equation}
Subject to the following constraints (equations of motion):
\begin{subequations}
\begin{align}
    \ddot{{\bf r}} = {\bf g} + \frac{{\bf T}}{m} \\
    \dot{m} = \frac{\|{\bf T}\|}{I_\text{sp}g_\text{ref}}
\end{align}
\end{subequations}
and the following boundary conditions:
\begin{subequations}
\begin{align}
    &{\bf r}(0) = {\bf r}_0 \\
    &{\bf v}(0) = \dot{{\bf r}}(0) = {\bf v}_0 \\
    &{\bf r}(t_f) = {\bf r}_{f} \\
    &{\bf v}(t_f) = \dot{{\bf r}}(t_f) = {\bf v}_{f}
\end{align}
\end{subequations}
And additional flight (glide slope) and thrust constraints:
\begin{subequations}
\begin{align}
     0 < T_{min} < \|{\bf T}\| < T_{max} \\
     \theta_{\text{alt}} = \arctan\left(\frac{\sqrt{r_y(t)^2+r_z(t)^2}}{r_x(t)}\right)<\tilde{\theta}_{alt}
\end{align}
\end{subequations}
Generally, one uses an inertial target-centered reference frame that neglects the rotation of Mars. Since the powered descent phase for a typical Mars mission is initiated at an altitude that is low w.r.t. the planet's radius, the gravity $\bf{g}$ is assumed to be constant (i.e., motion in a uniform gravitational field).  The lander's position is given by the vector ${\bf r} \in \mathbb{R}^3$ and ${\bf r}^T=[r_{x} \quad r_{y} \quad r_{z}]^T$, where $r_{x}$, $r_{y}$, and $r_{z}$ are the lander's downrange, crossrange, and altitude in the target centered reference frame. Likewise,the velocity is described by a vector ${\bf v} \in \mathbb{R}^3$ and ${\bf v}^T=[v_{x} \quad v_{y} \quad v_{z}]^T$, where $v_{x}$, $v_{y}$, and $v_{z}$ are the lander's downrange, crossrange, and descent components, respectively.  The vectors $\bf{r}_0, \bf{v}_0$, and $\bf{r}_f, \bf{v}_f$ are the initial and final position/velocity vetors, respectively. Importantly, the thrust magnitude $\|\bf{T}\|$ is limited between a maximum and minimum. The angle $\tilde{\theta}_\text{alt}<\frac{\pi}{2}$ represents that glide slope constraint. This is cast as a typical optimal control problem and generally solved using numerical methods (e.g. [\citenum{acikmese2007convex},\citenum{lu2017propellant}]). For a 6-DOF problem, one needs to provide additional information about the attitude of the lander.
The body frame is defined with the $z$-axis passing vertically through the lander and orthogonal $x$ and $y$ axes completing the body reference frame. Although a more formal 6-DOF trajectory and attitude optimal control problem can be done (e.g. [\citenum{szmuk2018successive}]), the RL framework is defined such that the agent (i.e., the lander) responds to a single cost (reward) signal which needs to be generally minimized (maximized) during the search process. Indeed, with the goal of the lander arriving at the origin of the target-centered reference frame at some specified velocity vector ${\bf v}=\dot{{\bf r}}$ and at a specified attitude, we can define the RL-based landing problem as follows:

\textbf{RL 6-DOF Closed-Loop Landing Problem}:

\noindent \textit{Minimize}:
\begin{enumerate}
    \item Terminal Position Error:  $\|{\bf r}\|$ at $t=t_{f}$
    \item Terminal Velocity Error:  $\|{\bf v}\|$ at $t=t_{f}$
    \item Terminal Attitude Error:  Magnitude of Pitch $\phi$ and Roll $\theta$ at $t=t_{f}$
    \item Terminal Rotational Velocity Error: $\|\bm{\omega}\|$ at $t=t_{f}$
    \item Control Effort: $\sum \| {\bf T}\|$ where the sum is over a trajectory and T is the total thrust
\end{enumerate}
\textit{Subject to}:
\begin{enumerate}
    \item Terminal Glideslope: $\|v_{z}\|/\|v_{x,y}\|>5$ over final 2 m of descent (soft constraint)
    \item Attitude Constraints: Magnitude of Pitch $\phi$ and Roll $\theta$ less than 80 degrees (hard constraint)
    \item Equations of Motion (set by the environment)
\end{enumerate}

Here, hard constraints are imposed by terminating the episode with a negative reward, whereas soft constraints are encouraged through rewards but without premature termination of the episode.
Next, the lander's equations of motion, the proposed policy gradient approach and a more detailed RL-based formulation for the cost function are presented.

\subsection{Equations of Motion}

The force ${\bf F}^{B}$ and torque ${\bf L}^{B}$ in the lander's body frame for a given commanded thrust depends on the placement of the thrusters in the lander structure. We can describe the placement of each thruster through a body-frame direction vector ${\bf {d}}$ and a position vector ${\bf r}$, both in $\mathbb{R}^3$. The direction vector is a unit vector giving the direction of the body frame force that results when the thruster is fired.  The position vector gives the body frame location where the force resulting from the thruster firing is applied for purposes of computing torque. For a lander with $k$ thrusters, the body frame force and torque associated with one or more  thrusters firing is given by, 
\begin{subequations}
\begin{align}
	{{\bf F}^{B}}&={\sum_{i=1}^{k}{\bf {d}}^{(i)} T_{cmd}^{(i)}}\label{eq:Thruster_modela}\\
	{{\bf L}^{B}}&={\sum_{i=1}^{k}{\bf r}^{(i)}\times{\bf F}_{B}^{(i)}}\label{eq:Thruster_modelb}
\end{align}
\end{subequations}
where $T_{cmd_{i}}\in[T_{min},T_{max}]$ is the commanded thrust for thruster $i$, $T_{min}$ and $T_{max}$ are a thruster's minimum and maximum thrust, ${\bf {d}}^{(i)}$ the direction vector for thruster $i$, and ${\bf r}^{(i)}$ the position of thruster $i$. The total body frame force and torque are calculated by summing the individual forces and torques.
The dynamics model uses the lander's current attitude ${\bf q}$ to convert the body frame thrust vector to the inertial frame is given by
\begin{equation}
	\label{eq:BtoN}
	{\bf F}^{N}=A^{N}_{B}({\bf q})^{T}{\bf F}^{B}
\end{equation}
where $A^{N}_{B}({\bf q})$ is the direction cosine matrix mapping the inertial frame to body frame obtained from the current attitude parameter ${\bf q}$. The attitude matrix is related to the quaternion by the following expression [\citenum{shuster:93}]: 
\begin{equation}
A^{N}_{B}({\bf q})=\Xi^T(\mathbf{q})\Psi(\mathbf{q})
\end{equation}
where
\begin{subequations}
\begin{gather}
\Xi(\mathbf{q})  \equiv \begin{bmatrix} q_4I_{3\times 3} +
[{\bm \varrho}\times] \\ -{\bm \varrho}^T\end{bmatrix} \\
\Psi(\mathbf{q})  \equiv\begin{bmatrix}q_4I_{3\times 3} - [{\bm
\varrho}\times] \\ -{\bm \varrho}^T \end{bmatrix}
\end{gather}
\end{subequations}
where the quaternion is divided into the $\bm \varrho$ and $q_4$, the vector and scalar components, respectively, and ${\bf q}^T=[{\bm \varrho} \quad q_4]^T$.The body force component can be rotated into the inertial frame using the following expression
\begin{equation}
	\label{eq:BtoN}
	{\bf F}^{N}=A^{N}_{B}({\bf q})^{T}{\bf F}^{B}
\end{equation}
The rotational velocities $\bm{\omega}_{B/N}$ are then obtained by integrating the Euler rotational equations of motion, is as follows [\citenum{junkins2009analytical}]:
\begin{equation}
	\label{eq:EulerRot}
	{\bf J}{\dot{\bm{\omega}}_{B/N}}=-\left[{\bm{\omega}}_{B/N}\times\right]{\bf J}\bm{\omega}_{B/N}+{\bf L}^{B}+{{\bf L}_\text{env}}^{B}
\end{equation}
with
\begin{equation}\label{cross}
[{\bf a} \times] \equiv \begin{bmatrix} 0 & -a_3 & a_2
\\ a_3 & 0 & -a_1 \\ -a_2 & a_1 & 0 \end{bmatrix}
\end{equation}
for any general $3 \times 1$ vector ${\bf a}$ defined such that
$[{\bf a}\times]{\bf b} = {\bf a}\times{\bf b}$.
The vector ${\bf L}^{B}$ is the body frame torque as given in Equation \eqref{eq:Thruster_modelb}, ${\bf L}_\text{env}$ is the body frame torque from external disturbances, and ${\bf J}$ is the lander's inertia tensor.
The lander's attitude is then updated by integrating the differential kinematic given by
\begin{equation}
    \label{eq:diffeqom}
 \dot{\bf{q}}=\frac{1}{2}\Xi\left({\bf q}\right){\bm \omega}_{B/N}
\end{equation}
where the lander's attitude is parameterized using the quaternion representation and $\bm{\omega}_{i}$ denotes the $i^{th}$ component of the rotational velocity vector $\bm{\omega}_{B/N}$.
The translational motion is modeled as follows
\begin{subequations}
\begin{align}
	{\Dot{\mathbf r}} &= {{\mathbf v}}\label{eq:EQOMa}\\
	{\Dot{\bf v}} &= {\frac{{{\bf F}^{N}}+{{\bf F}_\text{env}}^N}{m} + {\bf g}}\label{eq:EQOMb}\\
	\Dot{m} &= -\frac{\sum_{i}^{k}\lVert{{\bf F}^{B}}^{(i)}\rVert}{I_\text{sp}g_\text{ref}} \label{eq:EQOMc}
\end{align}
\end{subequations}
where ${{\bf F}^{N}}^{(i)}$ is the inertial frame force as given in Eq.~\eqref{eq:BtoN}, $k$ is the number of thrusters, $g_\text{ref}=9.8$ $\text{m}/\text{s}^{2}$,  ${\bf g}=\begin{bmatrix} 0 & 0 & -3.7114\end{bmatrix} \text{m}/\text{s}^2$ is used for Mars,  $I_\text{sp}=225$ s, and the spacecraft's mass is $m$.  ${\bf F}_\text{env}$ is a vector of normally distributed random variables representing environmental disturbances such as wind and variations in atmospheric density.
As a rough estimate of the force caused by wind gusts, a cross-sectional lander area of 100 square feet was assumed, and we used the Cornell University wind pressure calculator  to calculate the resulting wind pressure in pounds per square foot, divided this by 168 to account for the difference in average surface air pressure between the two planets, multiplied by the assumed cross-section of the lander, and then converted the force to Newtons. Using this rough approximation, a wind of 100 m/s would result in a force of 162 N. Note that the Mars GRAM model  shows a strong jet near 5-km altitude and 70 degrees N latitude with winds reaching 100 m/s, so it is reasonable to expect a Lander to be able to handle this case.  

For purposes of modeling the lander's moments of inertia, we model the lander as a uniform density ellipsoid, with inertia matrix given by
\begin{equation}
    \label{eq:inertia_tensor}
    {\bf J}=\frac{m}{5}\begin{bmatrix} b^2+c^2 & 0 & 0 \\ 0 & a^2+c^2 & 0 \\ 0 & 0 & a^2+b^2\end{bmatrix}
\end{equation}
where $a$, $b$, and $c$ correspond to the body frame $x$, $y$, and $z$ axes. $m$ is the lander's mass, which is updated as shown in Eq.~\eqref{eq:EQOMc}.
We assume the lander has a wet mass of 2000 kg and four throttleable thrusters with  a minimum and maximum thrust magnitude of 1000 N and 5000 N respectively. The four thrusters are located in the lander body frame as shown below in Table~\ref{tab:thrusters}, where $x$, $y$, and $z$ are the body frame axes. Roll is about the $x$-axis, yaw is about the $z$-axis, and pitch is about the $y$-axis. Note that this thruster configuration does not allow any direct control of the rotational velocity around the $z$-axis. However, the lander's yaw will change during the trajectory, but due to coupling with pitch via roll rather than due to torque caused by thrust.  Direct yaw control could be implemented by positioning the thrusters at a slight angle to the body-frame $z$-axis, which might result in faster convergence of policy optimization. The lander's translational motion is described using a target-centered inertial reference frame. The navigation system provides updates to the guidance system every 0.2 s, and we integrate the equations of motion using fourth order Runge-Kutta integration with a time step of 0.05 s.

\begin{table}[!ht]
	\fontsize{10}{10}\selectfont
    \caption{Body Frame Thruster Locations.}
   \label{tab:thrusters}
        \centering 
   \begin{tabular}{c | r | r | r } 
      \hline
      Thruster & x (m) & y (m) & z (m)\\
      \hline
      1 & 0 & -2 & -1 \\
      2 & 0 &  2 & -1 \\
      3 & -2 & 0 & -1 \\
      4 & 2 & 0 & -1 \\
   \end{tabular}
\end{table}

\subsection{Policy Gradient Method} \label{PGRL}
 In the RL framework, an agent learns through episodic interaction with an environment how to successfully complete a task by learning a \textit{policy} that maps observations to actions. The environment initializes an episode by randomly generating an internal state, mapping this internal state to an observation, and passing the observation to the agent. These observations could be a corrupted version of the internal state (to model sensor noise) or could be raw sensor outputs such as Doppler radar altimeter readings, or a multi-channel pixel map from an electro-optical sensor.  At each step of the episode, an observation is generated from the internal state and given to the agent. The agent uses this observation to generate an action that is sent to the environment; the environment then uses the action and the current state to generate the next state and a scalar reward signal.  The reward and the observation corresponding to the next state are then passed to the agent. The environment can terminate an episode, with the termination signaled to the agent via a done signal. The termination could be due to the agent completing the task or violating a constraint.  Initially, the agent's actions are random, which allows the agent to explore the state space and begin learning the value of experiencing a given observation, and which actions are to be preferred as a function of this observation. Here the value of an observation is the expected sum of discounted rewards received after experiencing that observation; this is similar to the cost-to-go in optimal control. As the agent gains experience, the amount of exploration is decreased, allowing the agent to exploit this experience. For most applications (unless a stochastic policy is required), when the policy is deployed in the field, exploration is turned off, as exploration gets quite expensive using an actual lander.  The safe method of continuous learning in the field is to have the lander send back telemetry data, which can be used to improve the environment's dynamics model and update the policy via simulated experience.
 
 In the following discussion, the vector ${\bf x}_k$ denotes the observation provided by the environment to the agent. Note that in general ${\bf x}_k$ does not need to satisfy the Markov property. In those cases where it does not, several techniques have proven successful in practice. In one approach, observations spanning multiple time steps  are concatenated, allowing the agent access to a short history of observations, which helps the agent infer the motion of objects in consecutive observations.  This was the approach used in [\citenum{mnih2015human}]. In another approach, a recurrent neural network is used for the policy and value function implementations.  The recurrent network allows the agent to infer motion from observations, similar to the way a recursive Bayesian filter can infer velocity from a history of position measurements. The use of recurrent network layers has proven effective in supervised learning tasks where a video stream needs to be mapped to a label [\citenum{baccouche2011sequential}].
 
 Each episode results in a trajectory defined by observation, actions, and rewards; a step in the trajectory at time $t_k$ can be represented as $({\bf x}_{k},{\bf u}_{k},r_{k})$, where ${\bf x}_k$ is the observation provided by the environment, ${\bf u}_k$ the action taken by the agent using the observation, and $r_k$ the reward returned by the environment to the agent.  The reward can be a function of both the observation ${\bf x}_{k}$ and the action ${\bf u}_k$. The reward is typically discounted to allow for infinite horizons and to facilitate temporal credit assignment. Then the sum of discounted rewards for a trajectory can be defined as
  \begin{equation}\label{objective_RL}
 r({\bm \tau})=\sum_{i=0}^{T}\gamma^i r_k({\bf x}_{k},{\bf u}_k)
  \end{equation}
  where ${\bm \tau}=[{\bf x}_{0},{\bf u}_0,...,{\bf x}_{T},{\bf u}_T]$ denotes the trajectory and $\gamma \in [0,1)$ denotes the discount factor.
 The objective function the RL methods seek to optimize is given by 
 \begin{equation}\label{objective_RL}
 J({\bm \theta})=\mathbb{E}_{p({\bm \tau})}\left[ r({\bm \tau})\right]=\int_{\mathbb{T}}r({\bm \tau})p_{\bm \theta}(\tau)d{\bm \tau}
 \end{equation}
where 
\begin{equation}
p_{\bm \theta}(\bm\tau)=\left[ \prod_{k=0}^{T}p({\bf x}_{k+1}|{\bf x}_k,{\bf u}_{\bm \theta})\right]p({\bf x}_0)
 \end{equation}
 where $\mathbb{E}_{p({\bm \tau})}\left[\cdot\right]$ denotes the expectation over trajectories and in general ${\bf u}_{\bm \theta}$ may be deterministic or stochastic function of the policy parameters, ${\bm \theta}$. However, it was noticed by Ref. \citenum{williams1992simple} that if the policy is chosen to be stochastic, where ${\bf u}_k\sim {\pi}_{\bm \theta}({\bf u}_k|{\bf x}_k)$ is a pdf for ${\bf u}_k$ conditioned on ${\bf x}_k$, then a simple policy gradient expression can be found. 
  \begin{equation}\label{policy_gradient_reinforce}
  \begin{aligned}
 \nabla_{\bm \theta} J({\bm \theta})=&\mathbb{E}_{p({\bm \tau})}\left[ \sum_{k=0}^{T}r_k({\bf x}_k,{\bf u}_k)  \nabla_{\bm \theta} \log  {\pi}_{\bm \theta}({\bf u}_k|{\bf x}_k) \right]\\
 &\approx \sum_{i=0}^{M}\sum_{k=0}^{T}r_k({\bf x}_k^i,{\bf u}_k^i)  \nabla_{\bm \theta} \log  {\pi}_{\bm \theta}({\bf u}_k^i|{\bf x}_k^i)
 \end{aligned}
 \end{equation}
 where the integral over ${\bm \tau}$ is approximated with $M$ samples from ${\bm \tau}^i\sim p_{\bm \theta}({\bm \tau})$ which are monte carlo roll-outs of the policy given the environment's transition pdf, $p({\bf x}_{k+1}|{\bf x}_k)$. The expression in Eq.~\eqref{policy_gradient_reinforce} is called the policy gradient and the form of this equation is referred to as the REINFORCE method [\citenum{williams1992simple}]. Since the development of the REINFORCE method additional theoretical work improved on the performance of the REINFORCE method. In particular, it was shown that the reward $r_k({\bf x}_k,{\bf u}_k)$ in Eq.~\eqref{policy_gradient_reinforce}  can be replaced with state-action value function $Q^{\pi}({\bf x}_k,{\bf u}_k)$, this result is known as the Policy Gradient Theorem. Furthermore, the variance of the policy gradient estimate that is derived from the monte carlo roll-outs, ${\bm \tau}^i$, is reduced by subtracting a state-dependent basis from $Q^{\pi}({\bf x}_k,{\bf u}_k)$. This basis is commonly chosen to be the state value function $V^{\pi}({\bf x}_k)$, and we can define $A^{\pi}({\bf x}_k,{\bf u}_k)=Q^{\pi}({\bf x}_k,{\bf u}_k)-V^{\pi}({\bf x}_k)$ . This method is known as the Advantage-Actor-Critic (A2C) Method. The policy gradient for the A2C method is given by
 \begin{equation}
\label{eq:pggrad}
\nabla_{{\bm \theta}}J({\bm \theta})\approx \sum_{i=0}^{M}\sum_{k=0}^{T}A^{\pi}_{\bf{w}}({\bf x}_{k}^i,{\bf u}_{k}^i)  \nabla_{\bm \theta} \log  {\pi}_{\bm \theta}({\bf u}_k^i|{\bf x}_k^i)
\end{equation}
where the  $\mathbf w$ subscript denotes a function parameterized by $\mathbf w$.

\subsection{Proximal Policy Optimization} \label{PGRL}
The Proximal Policy Optimization (PPO) approach  [\citenum{schulman2017proximal}] is a type of policy gradient which has demonstrated state-of-the-art performance for many RL benchmark problem. The PPO approach is developed using the properties of the Trust Region Policy Optimization (TRPO) Method [\citenum{schulman2015trust}]. The TRPO method formulates the policy optimization problem using a constraint to restrict the size of the gradient step taken during each iteration [\citenum{sorensen1982newton}]. The TRPO method policy update is calculated using the following problem statement:
\begin{equation}\label{TRPOeq}
\begin{aligned}
& \underset{{\bm \theta}}{\text{minimize}}
& & \mathbb{E}_{p({\bm \tau})}\left[\frac{\pi_{{\bm \theta}}({\bf u}_{k}|{\bf x}_{k})}{\pi_{{\bm \theta}_\text{old}}({\bf u}_{k}|{\bf x}_{k})}A^{\pi}_{\bf w}({\bf x}_{k},{\bf u}_{k})\right] \\
& \text{subject to}
& & \mathbb{E}_{p({\bm \tau})}\left[ \text{KL}\left( \pi_{{\bm \theta}}({\bf u}_{k}|{\bf x}_{k}),\pi_{{\bm \theta}_\text{old}}({\bf u}_{k}|{\bf x}_{k}) \right)  \right] \leqslant \delta
\end{aligned}
\end{equation}
where the function $\text{KL}(\cdot,\cdot)$ is the Kullback-Leibler (KL) divergence [\citenum{kullback1951information}]. The parameter $\delta$ is a tuning parameter but the theory justifying the TRPO methods proves monotonic improvement in the policy performance if the policy change in each iteration is bounded a parameter $C$. The parameter $C$ is computed using the Kullback-Leibler (KL) divergence [\citenum{kullback1951information}]. Reference \citenum{schulman2015trust} computes a closed-form expression for $C$ but this expression leads to prohibitively small steps, and therefore, Eq.~\eqref{TRPOeq} with a fix constraint is used. Additionally, Eq.~\eqref{TRPOeq} is approximately solved using the conjugate gradient algorithm, which approximates the constrained optimization problem given by Eq.~\eqref{TRPOeq} with a linearized objective function and a quadratic approximation for the constraint. The PPO method approximates the TRPO optimization process by accounting for the policy adjustment constrain with a clipped objective function. The objective function used with PPO can be expressed in terms of the probability ratio $p_{k}({\bm \theta})$ given by,
\begin{equation}
\label{eq:clipr}
p_{k}({\bm \theta})=\frac{\pi_{{\bm \theta}}({\bf u}_{k}|{\bf x}_{k})}{\pi_{{\bm \theta}_\text{old}}({\bf u}_{k}|{\bf x}_{k})}
\end{equation}
where the PPO objective function is then as follows:
\begin{equation}
\label{eq:ppoloss}
J({\bm \theta})=\mathbb{E}_{p({\bm \tau})}\left[\mathrm{min}\left[p_{k}({\bm \theta}) , \mathrm{clip}(p_{k}({\bm \theta}) , 1-\epsilon, 1+\epsilon)\right]A^{\pi}_{\bf w}({\bf x}_{k},{\bf u}_{k})\right]
\end{equation}
This clipped objective function has been shown to maintain the KL divergence constraints, which aids convergence by insuring that the policy does not change drastically between updates.
PPO uses an approximation to the advantage function that is the difference between the empirical return and a state value function baseline is given by the following:
\begin{equation}
\label{eq:ppo_adv}
	A^{\pi}_{\bf w}(\mathbf{x}_{k},\mathbf{u}_{k})=\left[\sum_{\ell=k}^{T}\gamma^{\ell-k}r(\bf x_{\ell},\bf u_{\ell})\right]-V_{\bf w}^{\pi}(\mathbf{x}_{k})
\end{equation}
Here the value function $V_{\bf w}^{\pi}$ is learned using the cost function given by
\begin{equation}
\label{eq:vf_ppo}
L(\mathbf{w})=\sum_{i=1}^{M}\left(V_{\mathbf{w}}^{\pi}({\bf x}_k^i)-\left[\sum_{\ell=k}^{T}\gamma^{\ell-k}r({\bf u}_{\ell}^i,{\bf x}_{\ell}^i)\right]\right)^2
\end{equation}
In practice, policy gradient algorithms update the policy using a batch of trajectories (roll-outs) collected by interaction with the environment. Each trajectory is associated with a single episode, with a sample from a trajectory collected at step $k$ consisting of observation ${\bf x}_{k}$, action ${\bf u}_{k}$, and reward $r_k({\bf x}_k,{\bf u}_k)$. Finally, gradient accent is performed on ${\bm \theta}$ and gradient decent on ${\bf w}$ and update equations are given by 
\begin{align}\label{loss}
{\bf w}^+&={\bf w}^--\beta_{{\bf w}}\nabla_{{\bf w}} \left. L({\bf w})\right|_{{\bf w}={\bf w}^-}\\
{\bm \theta}^+&={\bm \theta}^-+\beta_{{\bm \theta}} \left. \nabla_{\bm \theta}J\left({\bm \theta}\right)\right|_{{\bm \theta}={\bm \theta}^-}
\end{align}
where $\beta_{{\bf w}}$ and $\beta_{{\bm \theta}}$ are the learning rates for the value function, $V_{\bf w}^{\pi}\left({\bf x}_k\right)$, and policy, $\pi_{\bm \theta}\left({\bf u}_k|{\bf x}_k\right)$, respectively.
In our implementation, we adjust the clipping parameter $\epsilon$ to target a KL divergence between policy updates of 0.001. The policy and value function is learned concurrently, as the estimated value of a state is policy dependent. We use a Gaussian distribution with mean $\pi_{\bm \theta}({\bf u}_k|{\bf x}_{k})$ and a diagonal covariance matrix for the action distribution in the policy.  Because the log probabilities are calculated using the exploration variance, the degree of exploration automatically adapts during learning such that the objective function is maximized.

\subsection{Implementation Details}
In order to facilitate reproduction of our results, we include in this section several techniques we used in our implementation.  
We use the ADAM optimizer [\citenum{kingma2014adam}] to adjust the learning rate for both the policy and value function networks.
We use an approximation to the KL divergence for a Gaussian that is given by the mean square difference between the pre and post log probabilities of policy actions.  This approximation is close to the exact KL divergence, and is used in the open AI baseline PPO implementation PPO2. 
%
\begin{equation}
\label{eq:adj_kl}
\epsilon =
\begin{cases}
    \mathrm{min}(\epsilon_{max},1.5\epsilon) & \text{if } \mathrm{kl} < \frac{1}{2}\mathrm{kl_\text{targ}}   \\
    \mathrm{max}(\epsilon_{min},\frac{1}{1.5}\epsilon ) & \text{if } \mathrm{kl} > 2\mathrm{kl_\text{targ}} 
\end{cases}
\end{equation}

\begin{equation}
\label{eq:adj_lr}
\zeta =
\begin{cases}
    1.5\zeta & \text{if } \mathrm{kl} < \frac{1}{2}\mathrm{kl_\text{targ}}  \text{ and } \epsilon > \frac{1}{2}\epsilon_\text{max} \text{ and } \zeta < \zeta_\text{max} \\
    \frac{1}{1.5}\zeta & \text{if } \mathrm{kl} > 2\mathrm{kl_\text{targ}}  \text{ and } \epsilon < 2\epsilon_\text{min} \text{ and } \zeta > \zeta_\text{min}
\end{cases}
\end{equation}
We adjust the clipping parameter $\epsilon$ based off of the KL divergence between policy updates as shown in Equation \eqref{eq:adj_kl}, where kl is the KL divergence between policy updates, $\mathrm{kl_\text{targ}}$ is the target KL divergence, and $\epsilon_\text{max}$ and $\epsilon_\text{min}$ are set to 0.01 and 0.5.  A similar implementation was suggested in Reference [\citenum{schulman2017proximal}].  We also adjust the ADAM step size by multiplying it by a parameter $\zeta$ as shown in \eqref{eq:adj_lr}, where $\zeta_\text{min}$ and $\zeta_\text{max}$ are set to 0.1 and 10.  A learning rate adjustment was used in [\citenum{schulman2017proximal}] for the roboschool environments, but the exact implementation was not divulged. 

When approximating functions using artificial neural networks, it is important to scale the inputs to avoid saturating the activation functions at each layer. Most implementations scale the network input using statistics calculated over a given rollout.  Specifically, each element of an input vector is scaled by first subtracting the mean of the element and then dividing by three times the standard deviation, with the statistics calculated over a batch of rollouts. We use a variation on this approach that adjusts the statistics used for scaling incrementally over the entire optimization, which boosts performance by avoiding discontinuities in the scaling statistics.  It is also  important to insure that  the magnitude of the neural network outputs are are reasonably close to unity, although full normalization is not required. For policy parameter learning, we scale the action such that maximum thrust for an engine corresponds to a value of one.  For value function parameter learning, we use a heuristic that multiplies the rewards accumulated over an episode by a factor of $1-\gamma$.

\subsection{RL Problem Formulation}

In order to apply the reinforcement learning framework developed in Section \ref{PGRL} to a particular problem, we need to define an environment and reward function and specify the policy and value function network architectures. 
To test the policy, the trained policy is substituted for the agent. The action taken by the agent based on the observation provided by the dynamics model is interpreted as a thrust command by the thruster model, which passes a body frame force and torque to the dynamics model, which computes the next state. An episode terminates when the lander's altitude falls below zero or the attitude constraint is violated. The initial condition generator generates random initial conditions for the lander with values uniformly distributed between the minimum and the maximum values given in Table~\ref{tab:IC}. We used a reduced set of initial conditions in order to reduce simulation time to allow faster iteration over different approaches to solving the Mars landing problem using RL. 


\begin{table}[!ht]
	\fontsize{10}{10}\selectfont
    \caption{Lander Initial Conditions for Optimization.}
   \label{tab:IC}
        \centering 
   \newcolumntype{R}{>{\raggedleft\arraybackslash}p{1.8cm}}
   \begin{tabular}{l | R | R | R | R } 
      \hline 
       & \multicolumn{2}{c}{Velocity}\vline & \multicolumn{2}{c}{Position}\\
       \hline
       & min (m/s) & max (m/s) & min (m) & max (m) \\
       \hline
      Downrange      & -70 & -10 & 0 & 2000\\
      Crossrange       & -30  & 30 & -1000 & 1000 \\
      Elevation     & -90 & -70 & 2300 & 2400 \\
      \hline
      & min (rad/s) & max (rad/s) & min (rad) & max (rad) \\
      \hline
      Yaw  &  0.00 & 0.00 & $-\pi/8$ & $\pi/8$  \\
      pitch & -0.01 & 0.01 & $\pi/4-\pi/8$ & $\pi/4 + \pi/16$  \\
      roll & -0.01 & 0.01 & $-\pi/8$ & $\pi/8$  \\
   \end{tabular}
\end{table}

To speed up the design process, we developed a 3-DOF RL environment where we could quickly assess the impact of different reward functions and hyperparameter settings.  A policy optimized in the 3-DOF RL environment converges around 70 times faster than in the 6-DOF environment.  The  reward function for the 3-DOF environment is identical to that of the 6-DOF environment. We show optimization learning curves for the 3-DOF environment to allow comparison.

The policy and value functions are implemented using four-layer neural networks with tanh activations on each hidden layer.  The network architectures are as shown in Table \ref{tab:NN}, where $n_{\mathrm{hi}}$ is the number of units in layer $i$, $\mathrm{obs\_dim}$ is the observation dimension, and $\mathrm{act\_dim}$ is the action dimension.

\begin{table}[!ht]
	\fontsize{10}{10}\selectfont
    \caption{Policy and Value Function network architecture.}
   \label{tab:NN}
        \centering 
   \newcolumntype{R}{>{\raggedleft\arraybackslash}p{1.8cm}}
   \begin{tabular}{l | R | c | R | c } 
      \hline 
       & \multicolumn{2}{c}{Policy Network}\vline & \multicolumn{2}{c}{Value Network}\\
       \hline
       Layer & \# units & activation & \# units & activation \\
       \hline
      hidden 1      & $10 * \mathrm{obs\_dim}$ & tanh & $10 * \mathrm{obs\_dim}$ & tanh \\
      hidden 2      & $\sqrt{n_{\mathrm{h1}} * n_{\mathrm{h3}}}$ & tanh & $\sqrt{n_{\mathrm{h1}} * n_{\mathrm{h3}}}$ & tanh\\
      hidden 3      & $10 * \mathrm{act\_dim}$ & tanh & 5 & tanh \\
      output        & $\mathrm{act\_dim}$ & linear & 1 & linear \\
      \hline
   \end{tabular}
\end{table}

The most challenging part of solving the Mars landing problem using RL was the development of a reward function that works well in a sparse reward setting. If we only reward the agent for making a soft pinpoint landing at the correct attitude and with close to zero rotational velocity, the agent would never see the reward within a realistic number of episodes, as the probability of achieving such a landing using random actions in a 6-DOF environment with realistic initial conditions is exceedingly low. The sparse reward problem is typically addressed using inverse reinforcement learning [\citenum{ng2000algorithms}], where a per time step reward function is learned from expert demonstrations. With a reward given at each step of agent-environment interaction, the rewards are no longer sparse. In theory, demonstrations using optimal control could provide trajectories for inverse RL, but this would be very computationally expensive for 6-DOF trajectories, and many optimal control packages have trouble with complex non-convex and/or non-differentiable constraints.

Instead, we chose a different approach, where we engineer a reward function that, at each time step, provides hints to the agent (referred to as “shaping rewards”) that drive it towards a soft pinpoint landing. The recommended approach for such a reward shaping function is to make the reward a difference of potentials, in which case theoretical results have shown that the additional reward does not change the optimal policy [\citenum{ng2003shaping}].  We experimented with several potential functions with no success. Instead, we drew inspiration from biological systems that use the gaze heuristic. The gaze heuristic is used by animals such as hawks and cheetahs to intercept prey (and baseball players to catch fly balls) and works by keeping the line of sight angle constant during the intercept.  The gaze heuristic is also the basis of the well known PN guidance law used for homing phase missile guidance.  

In our case, the landing site is not maneuvering, and we have the additional constraint that we want the terminal velocity to be small. Therefore we use a heuristic where the agent attempts to keep its velocity vector aligned with the line of sight vector. Since the target is not moving in the target-centered reference frame, the target's future position is its current position, and the optimal action is to head directly towards the target. Such a rule results in a pinpoint, but not necessarily soft, landing.  To achieve the soft landing, the agent estimates time-to-go as the ratio of the range and the magnitude of the lander's velocity and reduces the targeted velocity as time-to-go decreases.  It is also important that the lander's terminal velocity be directed predominantly downward, the lander's terminal attitude is upright, and there are negligible terminal rotational velocity components. To achieve these requirements, we use the piecewise reward shaping function given below in Eqs.~\eqref{eq:vtarg1a}, \eqref{eq:vtarg1b}, \eqref{eq:vtarg1c}, \eqref{eq:vtarg1d}, and \eqref{eq:vtarg1e}, where $\tau_{1}$ and $\tau_{2}$ are hyperparameters  and $v_{o}$ is set to the magnitude of the lander's velocity at the start of the powered descent phase. We see that the shaping rewards take the form of a velocity field that maps the lander's position to a target velocity. In words, we target a location 15 m above the desired landing site and target a z-component of landing velocity equal to -2 m/s. Below 15 m, the downrange and crossrange velocity components of the target velocity field are set to zero, which encourages a vertical descent. Targeting a vertical descent has the beneficial side effect of encouraging the agent to keep the attitude level with no rotational velocity. This results in a good landing attitude with small rotational velocity components and a velocity directed primarily downward.  
\begin{subequations}
\begin{align}
    {\bf v}_{targ}&=-v_{o}\left(\frac{{\bf \hat{r}}}{\lVert{\bf \hat{r}}\rVert}\right)\left(1-\exp\left(-\frac{t_{go}}{\tau}\right)\right)\label{eq:vtarg1a}\\
    t_{go}&=\frac{\lVert{\bf \hat{r}}\rVert}{\lVert{\bf \hat{v}}\rVert}\label{eq:vtarg1b} \\
    {\bf \hat{r}}&=\begin{cases}{\bf r} - \begin{bmatrix}0 & 0 & 15\end{bmatrix}, & \text{if } r_{z} > 15 \\ \begin{bmatrix}0 & 0 & r_{z}\end{bmatrix}, &\text{otherwise}\end{cases}\label{eq:vtarg1c} \\
    {\bf \hat{v}}&=\begin{cases}{\bf v} - \begin{bmatrix}0 & 0 & -2\end{bmatrix}, & \text{if } r_{z} > 15 \\ {\bf v}-\begin{bmatrix}0 & 0 & -1\end{bmatrix}, &\text{otherwise}\end{cases}\label{eq:vtarg1d} \\
    \tau&=\begin{cases} \tau_{1}, & \text{if } r_{z} > 15 \\ \tau_{2}, &\text{otherwise}\end{cases}\label{eq:vtarg1e} 
\end{align}
\end{subequations}
Finally, we provide a terminal reward bonus when the lander reaches an altitude of zero, and the terminal position, velocity, attitude, and rotational velocity are within specified limits. The reward function is then given by the following:
\begin{equation}
 \begin{aligned}
 \label{eq:reward_func}
r &= \alpha\|{\bf v}-{\bf v}_{targ}\|+
\beta\|{\bf F}_{B}\|+ 
\gamma\mathrm{any}\left({\bf q}\left(t\right)>{\bf q}_{lim}\right)+
\delta\sum_{i=1}^{3}-\mathrm{max}(0,{\bf q}_{i}-{\bf q}_{mgn_{i}})  
+\eta\\
&+\kappa({ r}_{z}<0 \ \mathrm{and}\  
\|{\bf r}\|<r_{lim} \ \mathrm{and}\  \|{\bf v}\|<v_{lim} \ \mathrm{and}\ 
\mathrm{all({\bf q}<q_{lim}}) \ \mathrm{and}\ \mathrm{all(\boldsymbol{\omega}<\omega_{lim}}))
\end{aligned}
\end{equation}
where the various terms are described in the following:
\begin{enumerate}
    \item $\alpha$ weights a term penalizing the error in tracking the target velocity.
    \item $\beta$ weights a term penalizing control effort.
    \item $\gamma$ weights a term penalizing exceeding yaw, pitch, and roll limits. The "any" function is set to Boolean True if any elements of its argument are Boolean True (just like the Python np.any() function). If any attitude component exceeds its limit, the episode is terminated.
    \item $\delta$ weights a term that increases as the lander's attitude passes a threshold $q_{mgn}$; this gives the agent a hint that it is approaching the attitude limit $q_{lim}$
    \item $\eta$ is a constant positive term that encourages the agent to keep making progress along the trajectory. Since all other rewards are negative, without this term, an agent would be incentivized to violate the attitude constraint and prematurely terminate the episode to maximize the total discounted rewards received starting from the initial state.
    \item $\kappa$ is a bonus given for a successful landing, where terminal position, velocity, attitude, and rotational velocity are all within specified limits. The ``all" function is set to Boolean True if all elements of its argument are Boolean True (just like the Python np.all() function). The limits are $r_{lim}=5$ $\text{m}$, $v_{lim}=2$ $\text{m}$, $q_{lim}=0.2$ rad (except for yaw, which is not limited), and $w_{lim}=0.2$ rad/s.
\end{enumerate}
This reward function allows the agent to trade off between tracking the target velocity given in Eq.~\eqref{eq:vtarg1a}, conserving fuel, satisfying the attitude constraints, and maximizing the reward bonus given for a good landing.  Note that the constraints are not hard constraints such as might be imposed in an optimal control problem solved using collocation methods. However, the consequences of violating the constraints (a large negative reward and termination of the episode) are sufficient to ensure they are not violated once learning has converged. Hyperparameter settings and coefficients used in this work are given below in Table~\ref{tab:HPS}, note that due to lander symmetry, we do not impose any limits on the lander's yaw.
\begin{table}[!ht]
	\fontsize{10}{10}\selectfont
    \caption{Hyperparameter Settings.}
   \label{tab:HPS}
        \centering 
   \begin{tabular}{c | c | c | c | c | c | c | c | c | c | c } 
      \hline
      $v_{o}$ (m/s) & $\tau_{1}$ (s) & $\tau_{2}$ (s) & $\alpha$  & $\beta$   &  $\gamma$ & $\delta$ & $\eta$ & $\kappa$ & ${\bf q}_{lim}$ (rad) & ${\bf q}_{mgn}$ (rad)\\
      \hline
       $\|{\bf v}_{o}\|$ & 20 & 100 & -0.01    &  -0.05  & -100   &  -20  & 0.01 & 10 & $\begin{bmatrix}2\pi &\frac{7\pi}{16} &\frac{7\pi}{16}\end{bmatrix}$ & $\begin{bmatrix}0 & \frac{5\pi}{16} & \frac{5\pi}{16}\end{bmatrix}$\\
   \end{tabular}
\end{table}
The observation given to the agent during learning and testing is given by 
\begin{equation}
\label{eq:obs}
{\text{obs}}=\begin{bmatrix}{\bf v}_\text{error} &  {\bf q} & \bm{\omega} & r_{z} & t_\text{go}\end{bmatrix}
\end{equation}
where ${\bf v}_\text{error}={\bf v}-{\bf v}_\text{targ}$, with ${\bf v}_\text{targ}$ given in Eq.~\eqref{eq:vtarg1a}, the lander's estimated altitude, the time-to-go, as well as an estimate of the lander's attitude (${\bf q}$) and rotational velocity (${\bm \omega}$). Note that aside from the altitude, the lander translational coordinates do not appear in the observation. This results in a policy with good generalization in that the policy's behavior can extend to areas of the full state space that were not experienced during learning.
To provide a robust final policy, we optimize with parameter uncertainty.  Specifically, at the beginning of each episode both the lander's initial mass and the acceleration due to gravity are perturbed to give a random value within 5\% and 2\% of nominal, respectively.  In addition, we apply random uniform noise to the inertial tensor at the beginning of each episode. There is no physical significance to the inertia tensor noise; it is just a method of introducing parameter uncertainty in order to learn a robust policy.

\begin{table}[!ht]
	\fontsize{10}{10}\selectfont
    \caption{Parameter Uncertainty for Policy Optimization.}
   \label{tab:PUPO}
        \centering 
   \newcolumntype{R}{>{\raggedleft\arraybackslash}p{3.5cm}}
   \newcolumntype{C}{>{\center\arraybackslash}p{3.5cm}}
    \begin{tabular}{l | R | R }
    & min & max \\
       \hline
      Initial Mass (kg)    & 1900 & 2100 \\
      Gravitational Acceleration ($\mathrm{m/s}^2$)  &  $\begin{bmatrix}0.07 & 0.07 & 3.64 \end{bmatrix}$   &  $\begin{bmatrix}-0.07 & -0.07 & 3.79 \end{bmatrix}$\\
      Inertia Tensor Diagonal Noise ($\mathrm{kg}$-$\mathrm{m}^2$) & -100 & 100\\
      Inertia Tensor Off-Diagonal Noise ($\mathrm{kg}$-$\mathrm{m}^2$) & -10 & 10\\
   \end{tabular}
\end{table}

It turns out that the when a terminal reward is used (as we do), it is advantageous to use a relatively large discount rate.  However, it is also advantageous to use a lower discount rate for the shaping rewards.  To our knowledge, a method of resolving this conflict has not been reported in the literature.  In this work, we resolve the conflict by introducing a framework for accommodating multiple discount rates.  Let $\gamma_{1}$ be the discount rate used to discount $r_{1}(k)$, the reward function term (as given in \ref{eq:reward_func})  associated with the $\kappa$ coefficient). Moreover, let $\gamma_{2}$ be the discount rate used to discount $r_{2}(k)$, the sum of all other terms in the reward function. We can then rewrite the Eq.~\eqref{eq:vf_ppo} and \eqref{eq:ppo_adv} in terms of these rewards and discount rates, as given by the following:
\begin{subequations}
\begin{align}
J({\mathbf{w}})&=\sum_{i=1}^M \left(V_{\bf w}^{\pi}({\bf x}_{k}^i)-\left[\sum_{\tau=k}^{T}\gamma_{1}^{\ell-k}r_{1}({\bf u}_{\ell}^i,{\bf x}_{\ell}^i)+\gamma_{2}^{\ell-k}r_{2}({\bf u}_{\ell}^i,{\bf x}_{\ell}^i)\right]\right)^2\label{eq:new_vf}\\
A^{\pi}_{\mathbf{w}}({\bf x}_{k},{\bf u}_{k})&=\left[\sum_{\ell=k}^{T}\gamma_{1}^{\ell-k}r_{1}({\bf u}_{\ell},{\bf x}_{\ell})+\gamma_{2}^{\ell-k}r_{2}({\bf u}_{\ell},{\bf x}_{\ell})\right]-V_{\mathbf{w}}^{\pi}({\bf x}_{k})\label{eq:new_adv}
\end{align}
\end{subequations}
Although the approach is simple, the performance improvement is significant, although we had to give up the use of generalized advantage estimation [\citenum{schulman2015high}] as it is not compatible with multiple discount rates.  Without the use of multiple discount rates, the performance was actually worsened by including the terminal reward term. 

\section{Results}

\subsection{Policy Optimization}

\begin{figure}[!ht]
\begin{center}
\psfrag{Episode}[][]{{Episode}}
\psfrag{KL}[][]{{KL}}
\psfrag{Entropy, Explained Var}[][]{{Entropy, Explained Var}}
\psfrag{Entropy}[][]{\scriptsize{Entropy}}
\psfrag{Explained Var}[][]{\scriptsize{Explained V.}}
\psfrag{KL Divergence}[][]{{KL Divergence}}
\includegraphics[width=.8\linewidth]{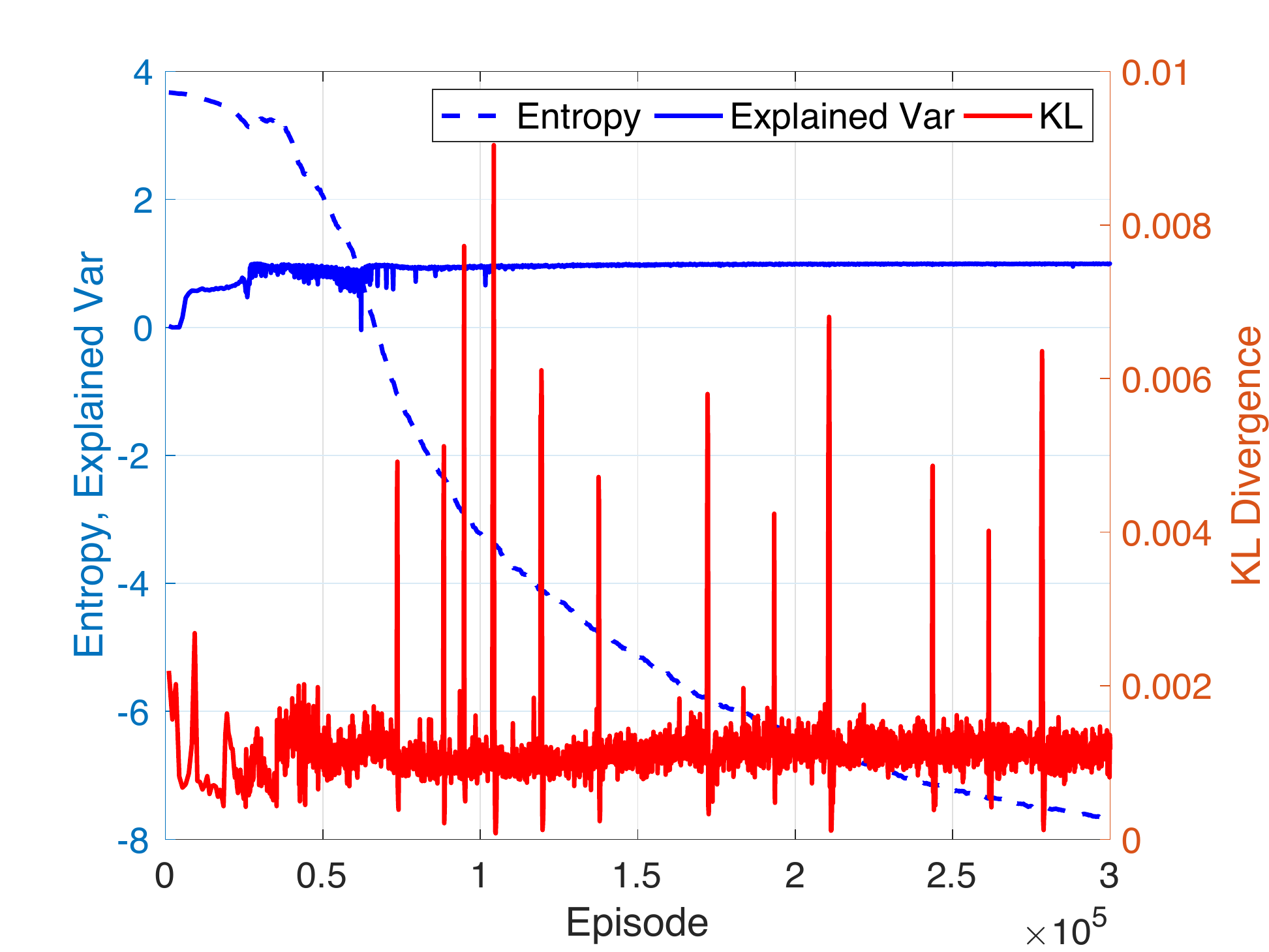}
\caption{6-DOF Optimization: Learning Statistics.}
\label{fig:lc2}
\end{center}
\end{figure}

\begin{figure*}[!ht]
\begin{centering}
\subfigure[Norm Position]{
\psfrag{Norm Position (m)}[][]{\scriptsize{Position (m)}}
\psfrag{Episode}[][]{\scriptsize{Episode}}
\psfrag{Terminal}[][]{\scriptsize{Terminal}}
\psfrag{Std}[][]{\scriptsize{Std}}
\includegraphics[keepaspectratio, width=0.45\textwidth]{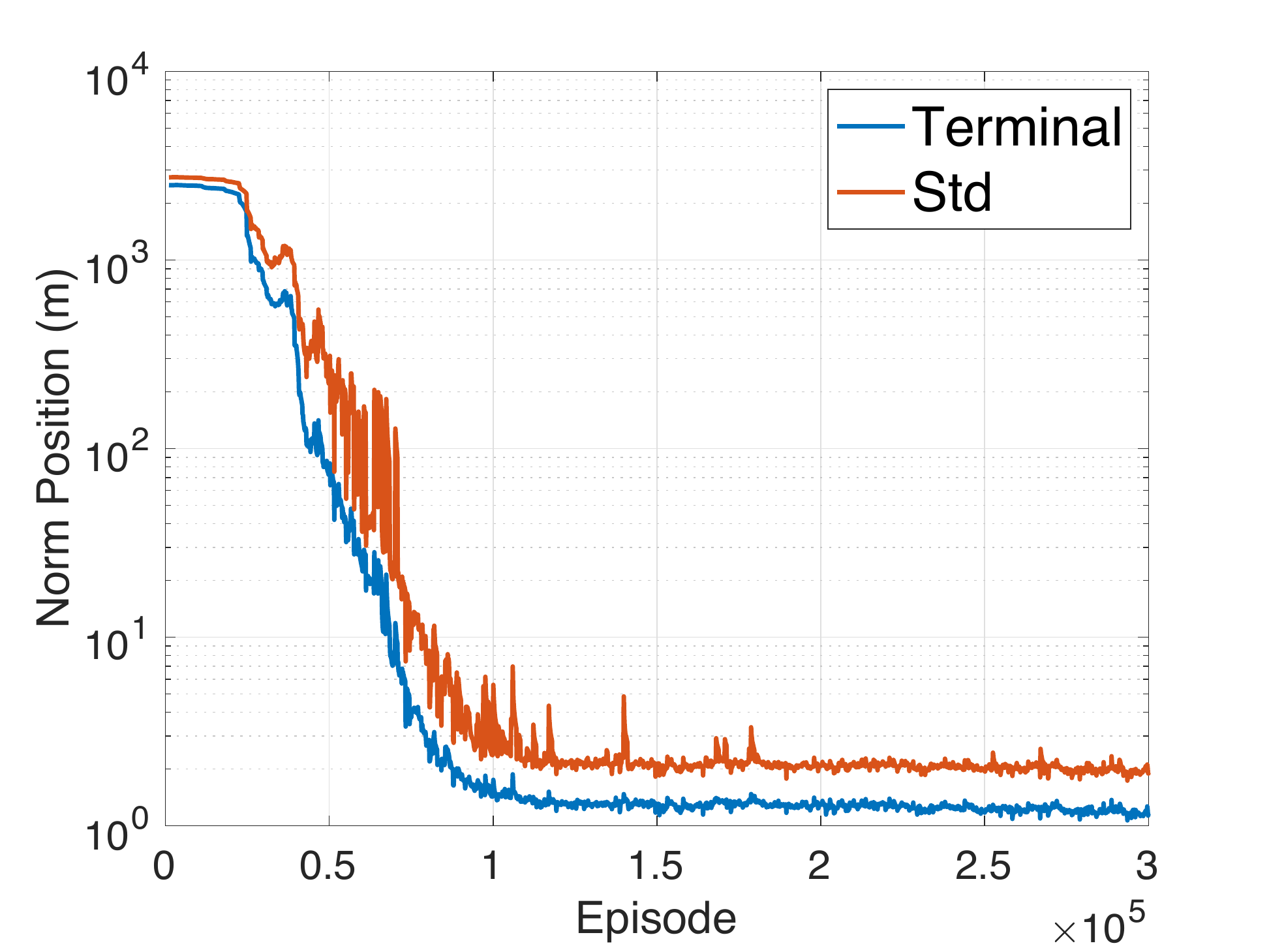}\label{fig:lc3}}
\subfigure[Norm Velocity]{
\psfrag{Norm Velocity (m/s)}[][]{\scriptsize{Velocity (m/s)}}
\psfrag{Std}[][]{\scriptsize{Std}}
\psfrag{Episode}[][]{\scriptsize{Episode}}
\psfrag{Terminal}[][]{\scriptsize{Terminal}}
\includegraphics[keepaspectratio, width=0.45\textwidth]{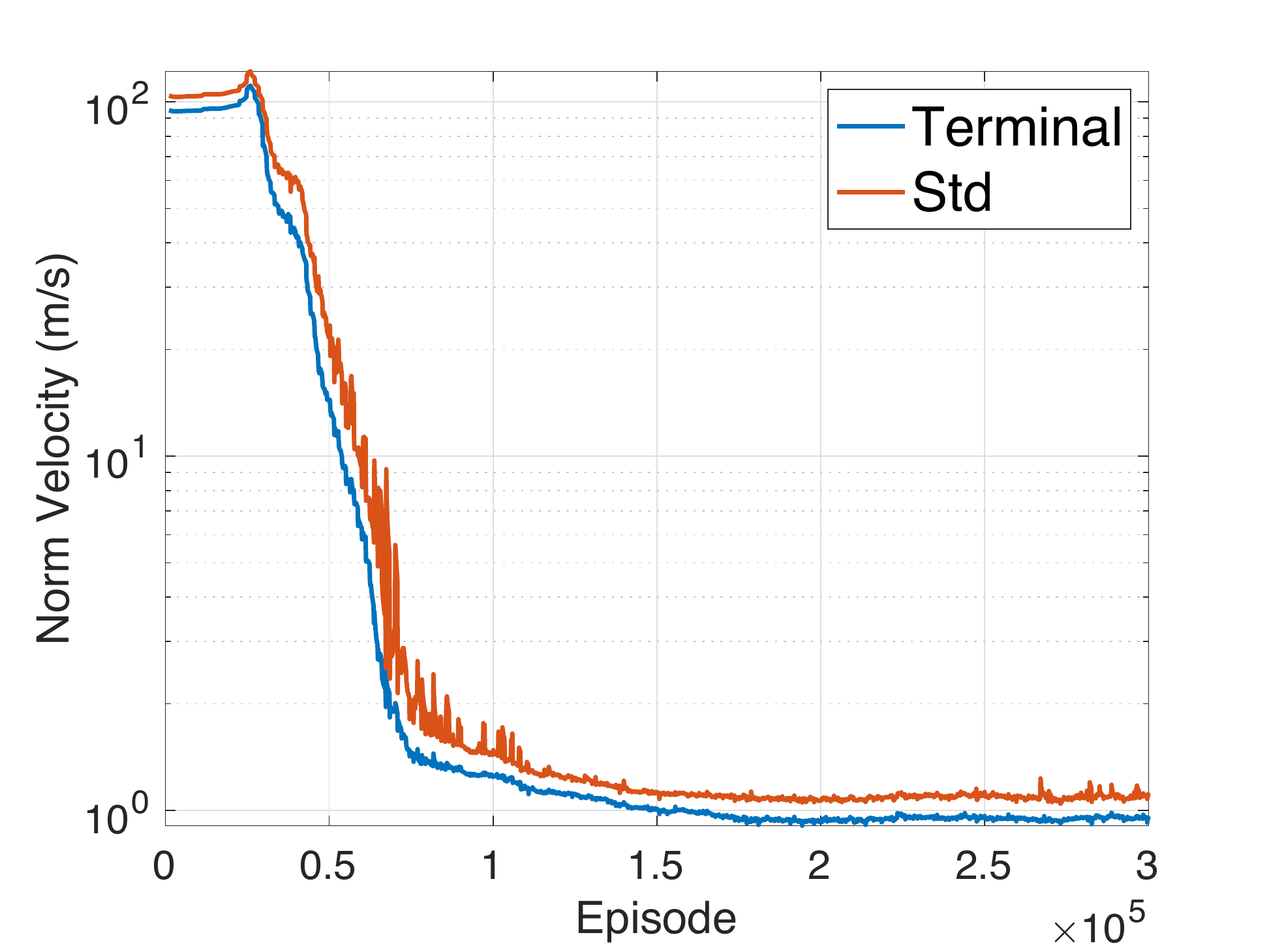}\label{fig:lc4}}
\\
\subfigure[Norm Attitude]{
\psfrag{Norm Attitude (Deg)}[][]{\scriptsize{Attitude (Deg)}}
\psfrag{Episode}[][]{\scriptsize{Episode}}
\psfrag{Std}[][]{\scriptsize{Std}}
\psfrag{Terminal}[][]{\scriptsize{Terminal}}
\includegraphics[keepaspectratio, width=0.45\textwidth]{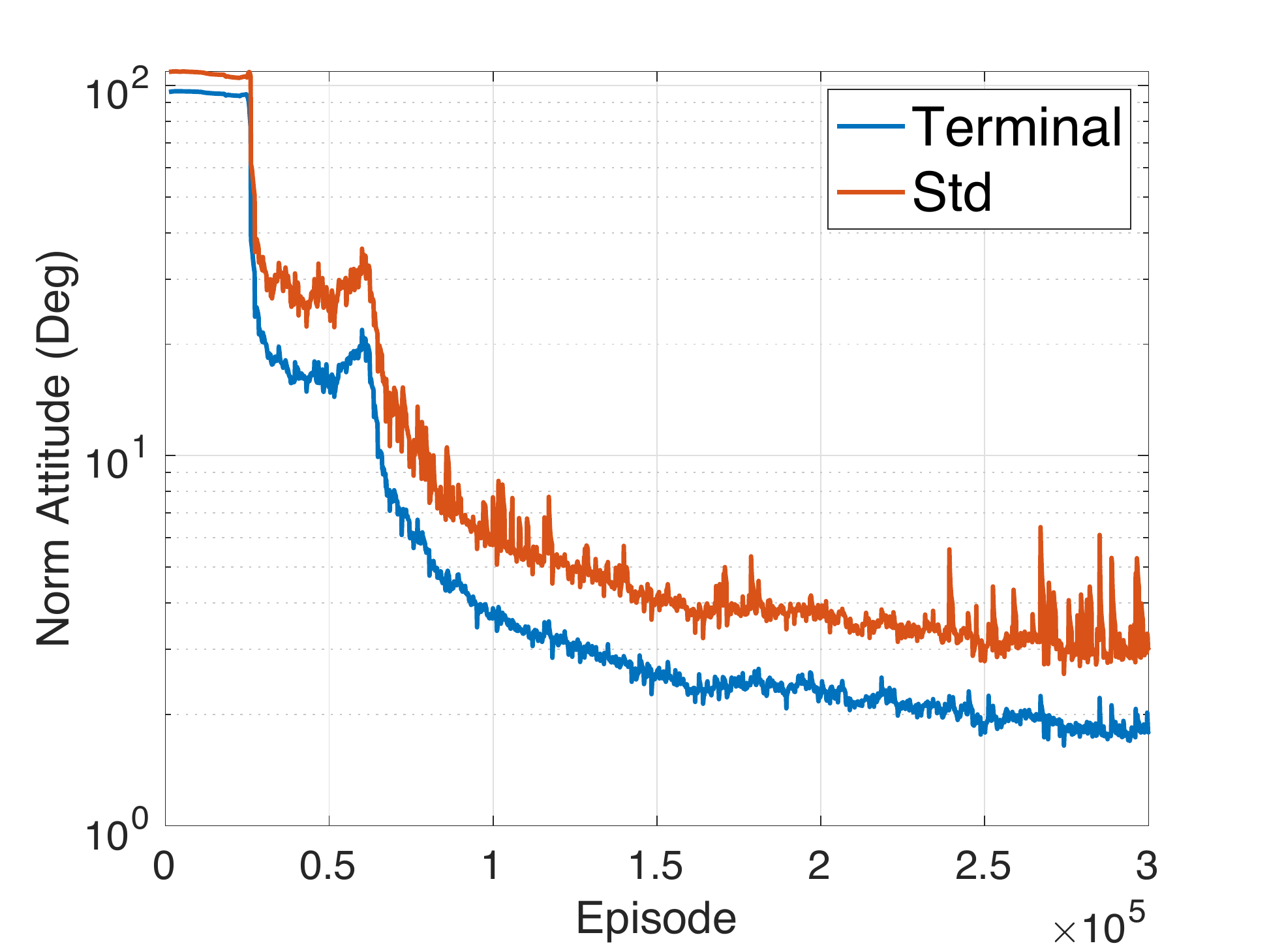}\label{fig:att}}
\subfigure[Norm Angular Velocity]{
\psfrag{Norm Angular Velocity (Deg/s)}[][]{\scriptsize{Angular Velocity (Deg/s)}}
\psfrag{Episode}[][]{\scriptsize{Episode}}
\psfrag{Std}[][]{\scriptsize{Std}}
\psfrag{Terminal}[][]{\scriptsize{Terminal}}
\includegraphics[keepaspectratio, width=0.45\textwidth]{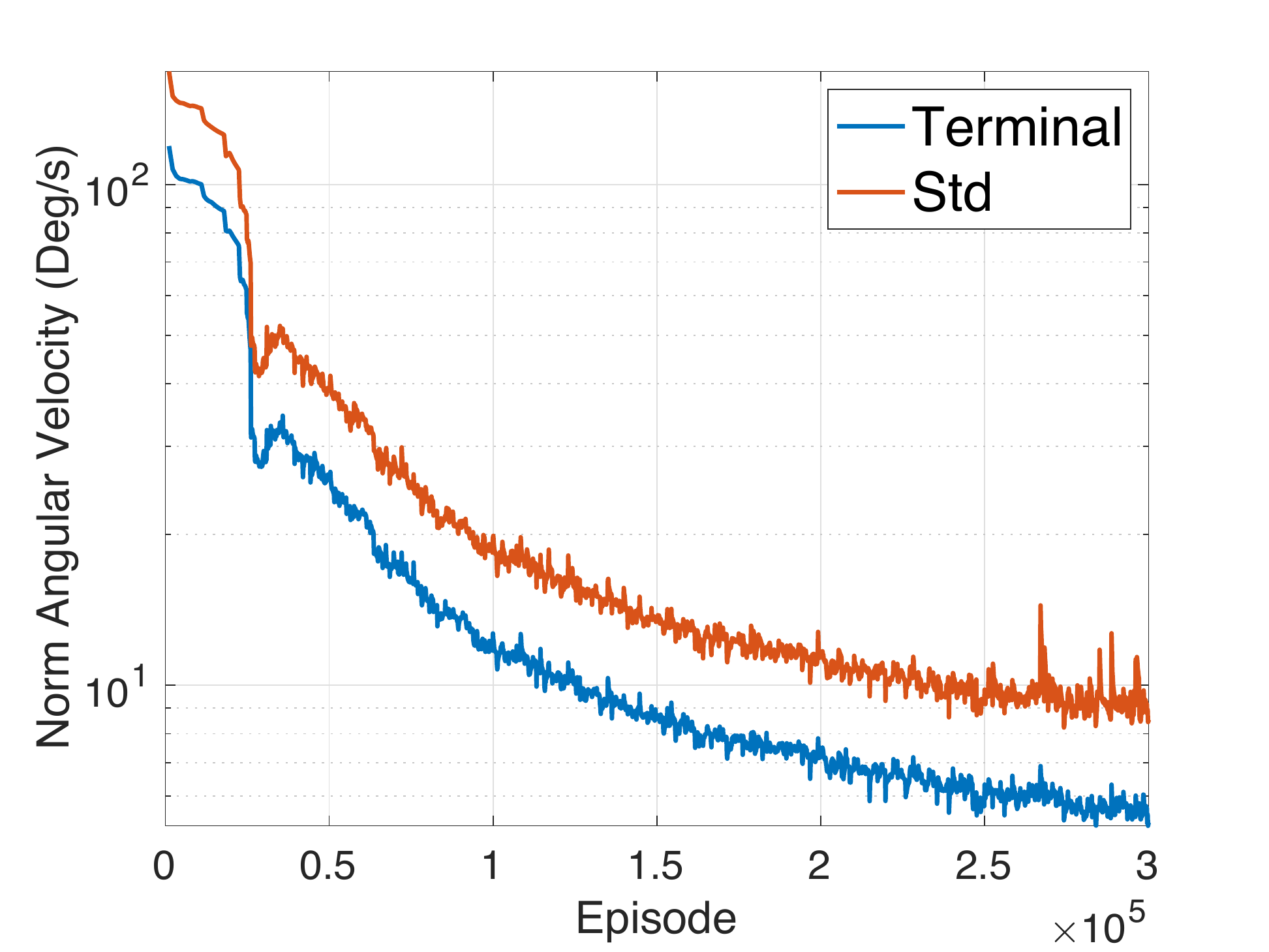}\label{fig:omg}}
\caption{Optimization: Lander Position, Velocity, Attitude, and Angular Velocity at end of Episode.}\label{Terminal}
\end{centering}
\end{figure*}

\begin{figure}[!ht]
\begin{center}
\psfrag{Episode}[][]{{Episode}}
\psfrag{KL}[][]{{KL}}
\psfrag{Entropy, Explained Var}[][]{{Entropy, Explained Var}}
\psfrag{Entropy}[][]{\scriptsize{Entropy}}
\psfrag{Explained Var}[][]{\scriptsize{Explained V.}}
\psfrag{KL Divergence}[][]{{KL Divergence}}
\includegraphics[width=.8\linewidth]{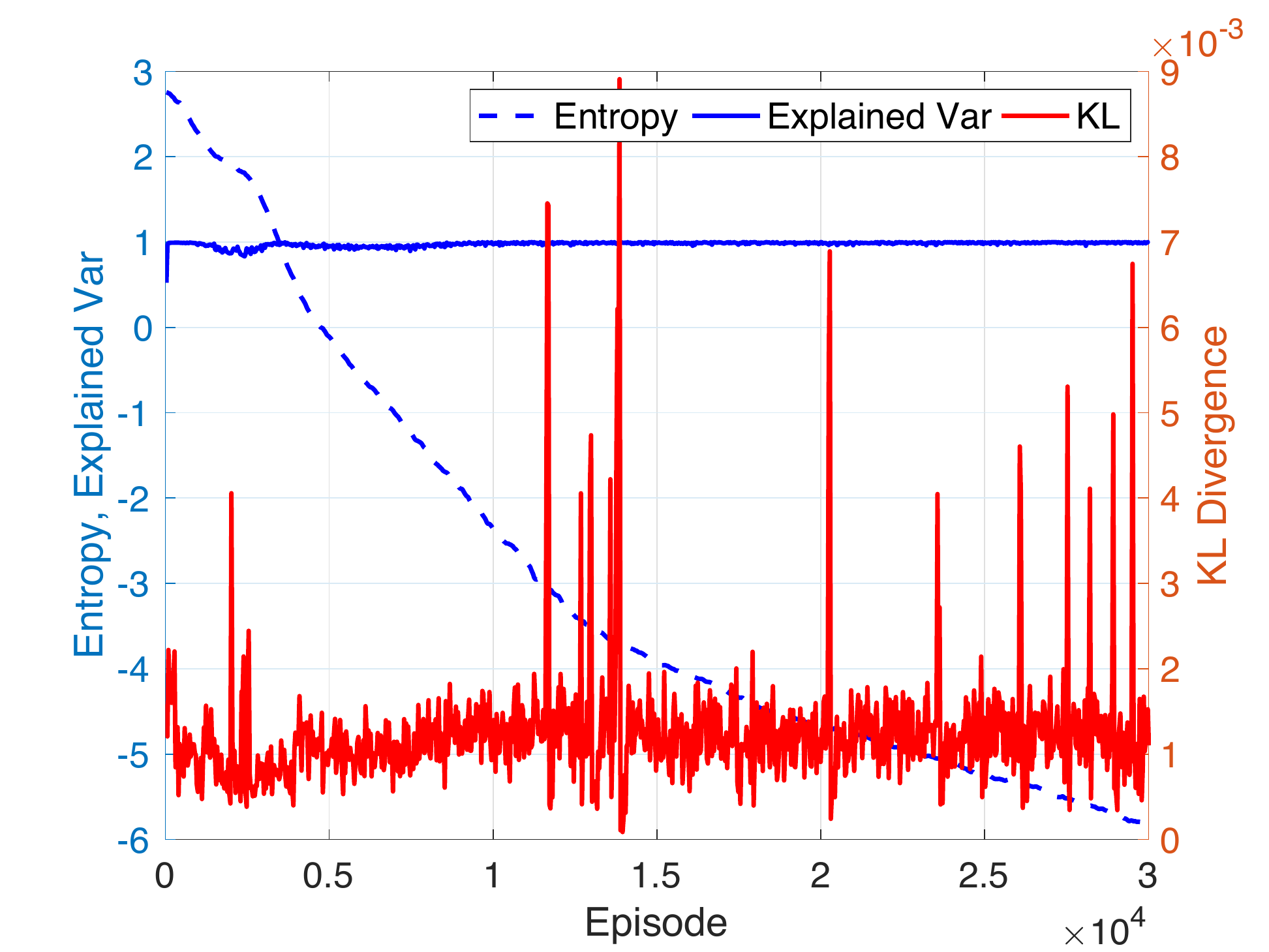}
\caption{3-DOF Optimization: Learning Statistics.}
\label{fig:3dof_statistics}
\end{center}
\end{figure}

\begin{figure*}[!ht]
\begin{centering}
\subfigure[Norm Position]{
\psfrag{Norm Position (m)}[][]{\scriptsize{Position (m)}}
\psfrag{Episode}[][]{\scriptsize{Episode}}
\psfrag{Terminal}[][]{\scriptsize{Terminal}}
\psfrag{Std}[][]{\scriptsize{Std}}
\includegraphics[keepaspectratio, width=0.45\textwidth]{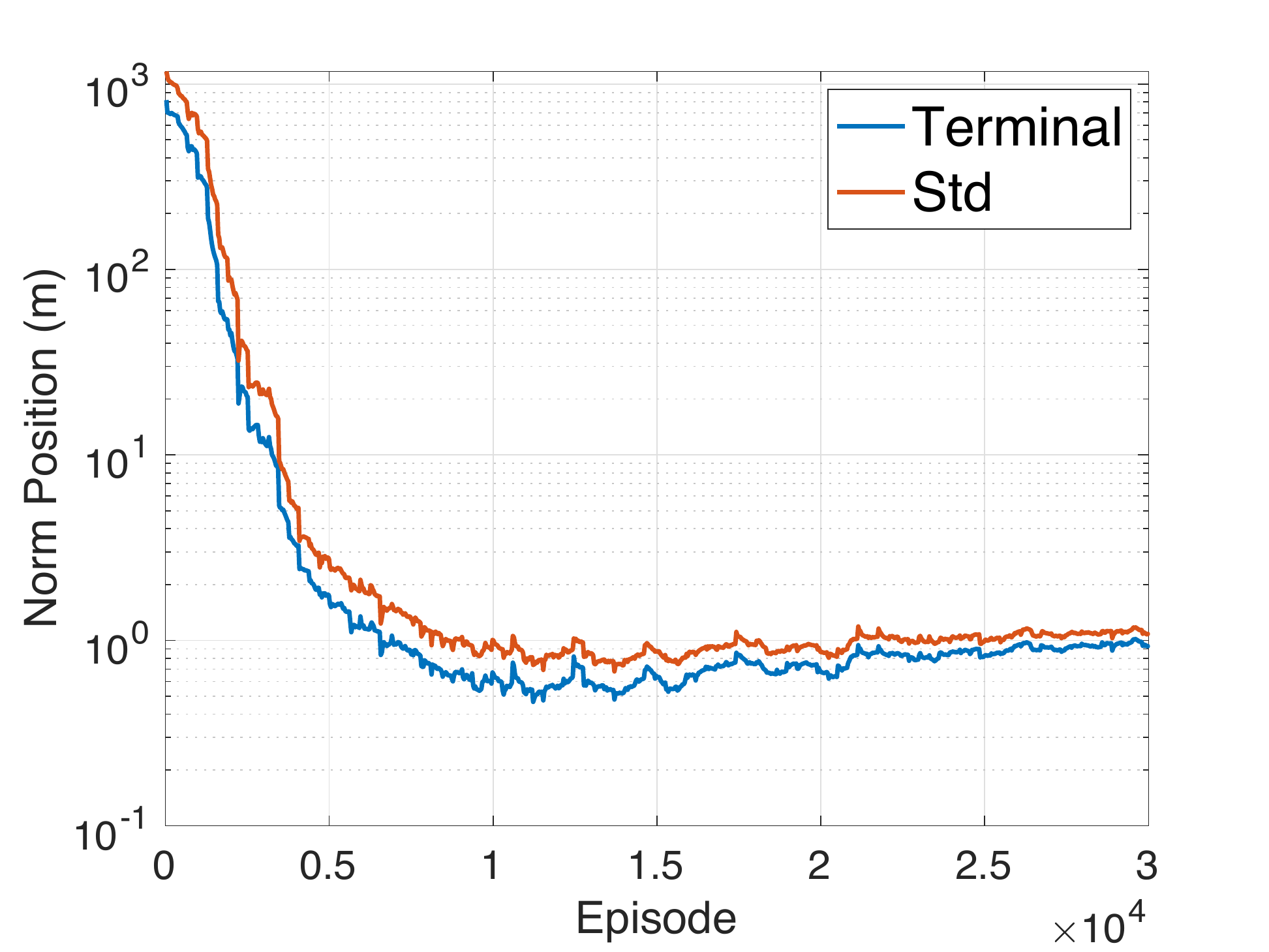}\label{fig:lc_3dof_rf}}
\subfigure[Norm Velocity]{
\psfrag{Norm Velocity (m/s)}[][]{\scriptsize{Velocity (m/s)}}
\psfrag{Std}[][]{\scriptsize{Std}}
\psfrag{Episode}[][]{\scriptsize{Episode}}
\psfrag{Terminal}[][]{\scriptsize{Terminal}}
\includegraphics[keepaspectratio, width=0.45\textwidth]{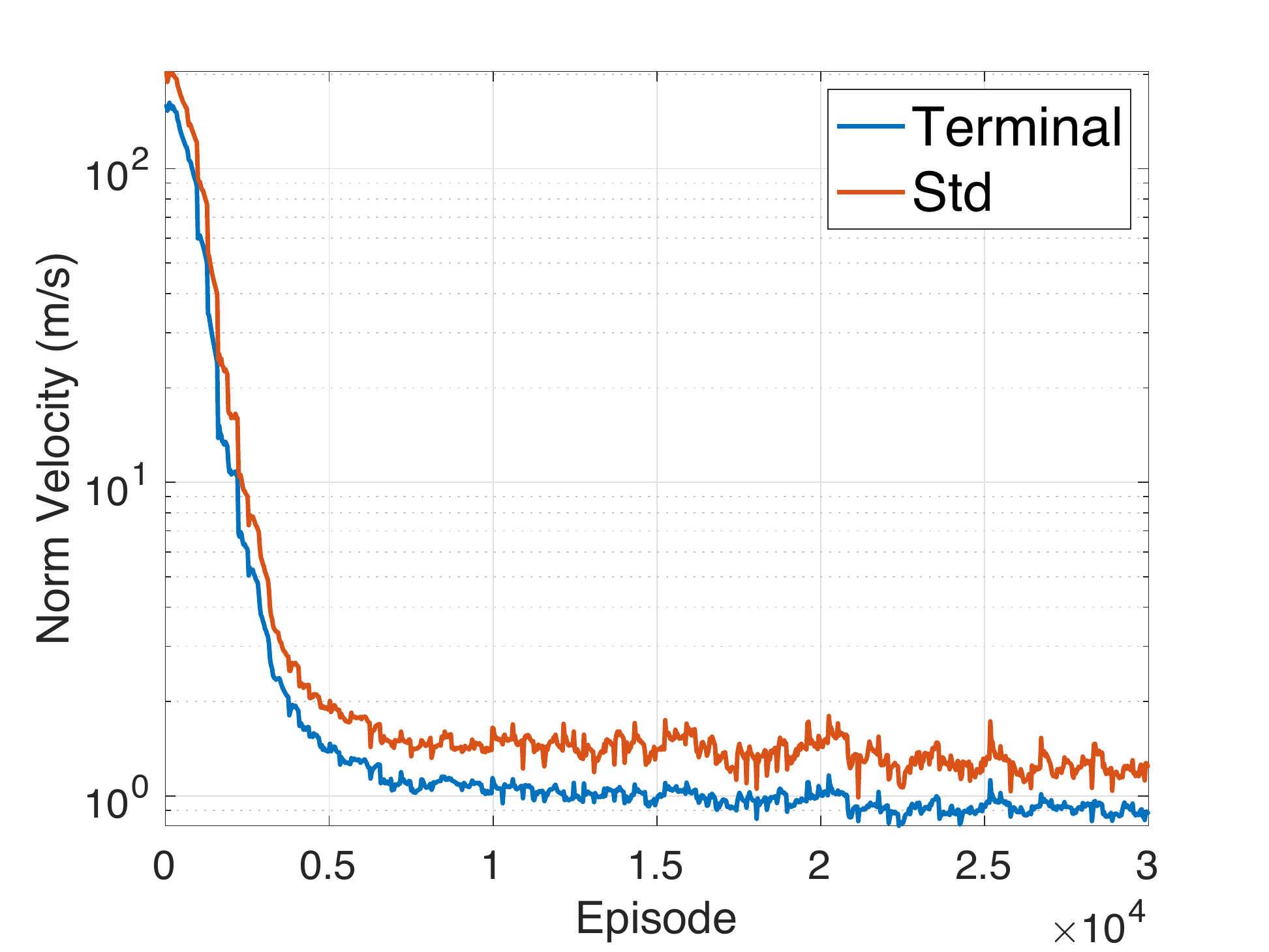}\label{fig:lc_3dof_vf}}
\caption{3-DOF Optimization: Lander Position, Velocity, Attitude, and Angular Velocity at end of Episode.}\label{Terminal_3dof}
\end{centering}
\end{figure*}

 Rollouts are generated by the agent interacting with the environment for 120 episodes, with the resulting trajectories used to compute the advantages and update the value and policy function approximators. Optimization was terminated after 300,000 episodes. Learning curves are shown below. In Figure \ref{fig:lc1} (Figure \ref{fig:3dof_rewards} for 3-DOF optimization), we plot statistics for the undiscounted rewards over 120 episodes, which is the number of episodes the agent accumulates before updating the policy and value function. "Steps" refers to the number of interactions between the agent and environment for the episode; this can be converted to the trajectory duration by multiplying by the navigation period of 0.2s. Figure \ref{fig:lc2} (Figure \ref{fig:3dof_statistics} for 3-DOF optimization) gives the policy entropy and the KL divergence between policy updates. We also plot the explained variance as a measure of how well the value function explains the observed returns; if the explained variance is 1, the value function perfectly explains the observed returns, if it is less than zero, you would have been better off predicting a constant value for the value of being in any state.  Figure \ref{fig:lc3} and Figure \ref{fig:lc4} (Figures \ref{fig:lc_3dof_rf} and \ref{fig:lc_3dof_vf} for 3-DOF) plot the lander's end of episode position and velocity magnitudes as learning progresses, where again the statistics are accumulated over the 120 episodes used for the policy and value function updates. Note in Figures \ref{fig:att} through \ref{fig:omg} that the standard deviation of the attitude and rotational velocity seems inconsistent with a good landing.  This is partly due to the parameter uncertainty introduced during learning, and partly due to policy exploration.   Although we terminated optimization at 300,000 episodes, the performance was still improving, and the standard deviation of the action was still around 4\% for exploration. Since this is 4\% of the maximum thrust, exploration noise still had a significant impact on the policy. During policy testing, we see that the magnitude of the attitude and rotational velocity is bounded more tightly.


\begin{figure}[!ht]
\begin{center}
\psfrag{Episode}[][]{{Episode}}
\psfrag{Steps}[][]{{Steps}}
\psfrag{Reward}[][]{{Reward}}
\psfrag{Mean Steps}[][]{\scriptsize{Mean Steps}}
\psfrag{Mean Reward}[][]{\scriptsize{Mean Reward}}
\psfrag{Std Reward}[][]{\scriptsize{Std Reward}}
\psfrag{MinReward}[][]{\scriptsize{Min Reward}}
\includegraphics[width=.8\linewidth]{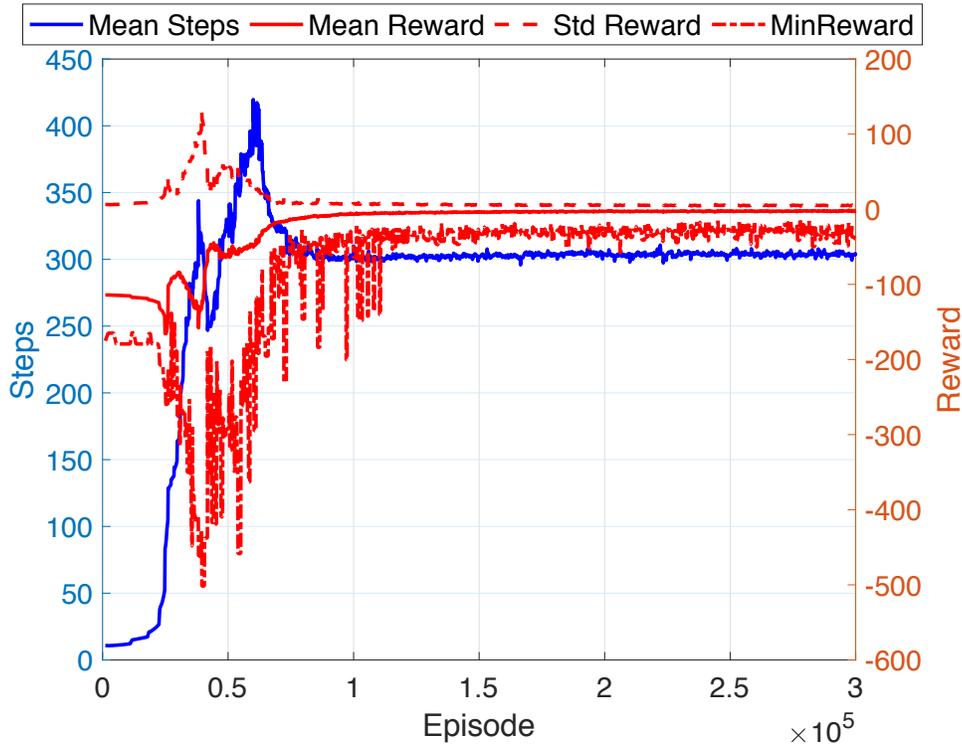}
\caption{6-DOF Optimization: Learning Curves.}
\label{fig:lc1}
\end{center}
\end{figure}


\begin{figure}[!ht]
\begin{center}
\psfrag{Episode}[][]{{Episode}}
\psfrag{Steps}[][]{{Steps}}
\psfrag{Reward}[][]{{Reward}}
\psfrag{Mean Steps}[][]{\scriptsize{Mean Steps}}
\psfrag{Mean Reward}[][]{\scriptsize{Mean Reward}}
\psfrag{Std Reward}[][]{\scriptsize{Std Reward}}
\psfrag{MinReward}[][]{\scriptsize{Min Reward}}
\includegraphics[width=.8\linewidth]{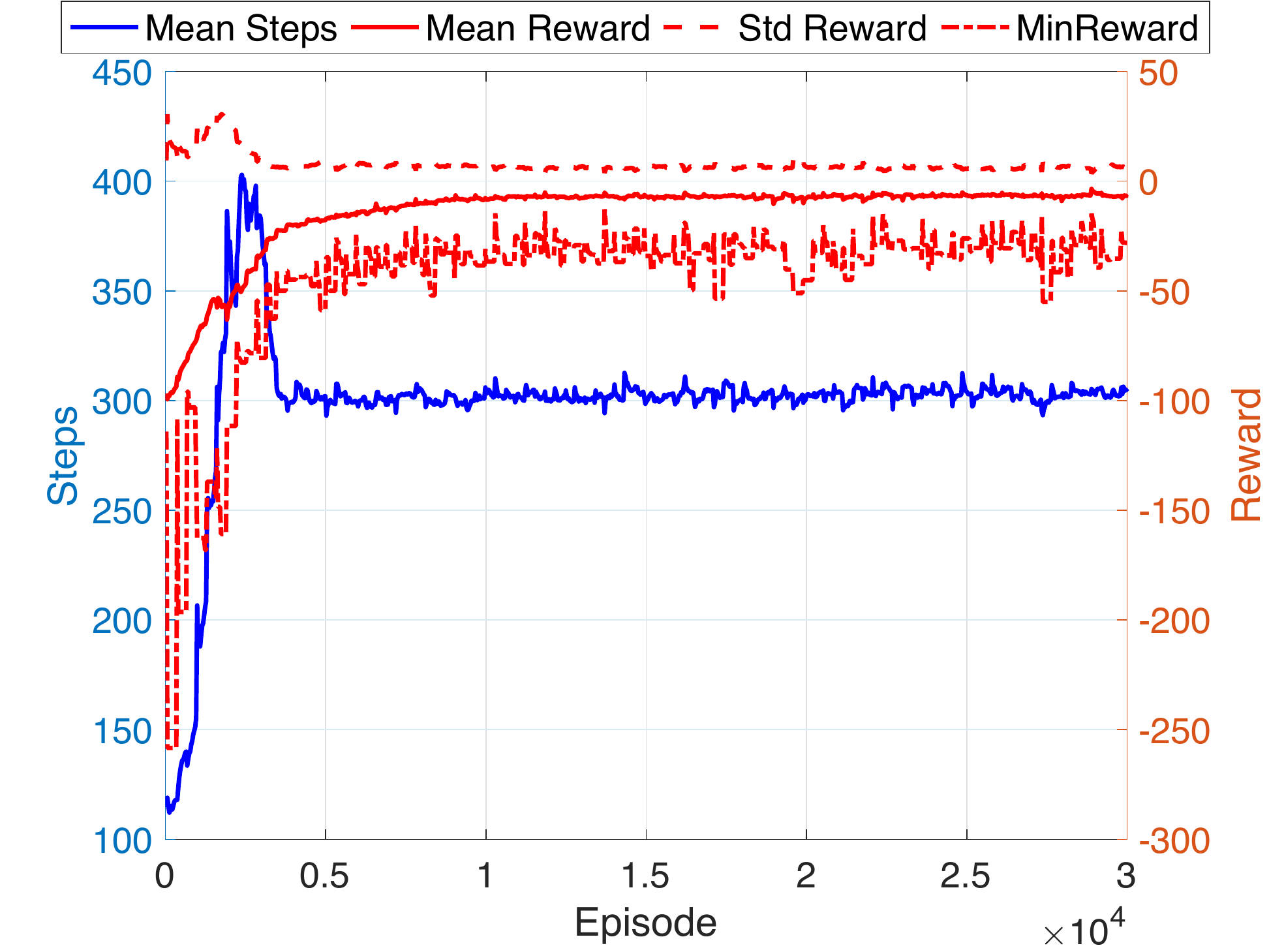}
\caption{3-DOF Optimization: Learning Curves.}
\label{fig:3dof_rewards}
\end{center}
\end{figure}

\subsection{Policy Testing}
To test the guidance policy, we simulate the policy for 10000 episodes using the same initial conditions as used for policy optimization, as given in Table~\ref{tab:IC}.  During testing, the dynamics model adds Gaussian noise to the force acting on the lander.  The noise has a mean that is computed at the start of each episode uniformly distributed between -100 and 100 N and held constant during the episode.  The noise standard deviation is 100 N. At the start of each episode, the nominal wet mass of 2000 kg is set to a uniformly distributed value between +/- 5\% of nominal. Table~\ref{tab:TDS} tabulates the touchdown statistics accumulated over testing. The glideslope statistic is given by the average $v_{z} / \lVert[{v}_{x},{v}_{y}]\rVert$ over the final 2 m of descent; here a high value is desirable, as it indicates the lander's velocity is directed primarily in the downward direction.  Note that due to symmetry in the lander design, we do not care about the terminal yaw value.

\begin{table}[!h]
	\fontsize{10}{10}\selectfont
    \caption{Touchdown Statistics (same initial condition range as for optimization).}
   \label{tab:TDS}
   \centering 
   \newcolumntype{R}{>{\raggedleft\arraybackslash}p{1.5cm}}

   \begin{tabular}{l | R | R | R | R } 
      \hline
        & Mean  & Std. Dev.  &  Min & Max \\
      \hline
       Downrange Position  (m)   & 0.4  &  0.7 & -2.9 & 4.5 \\
       Crossrange Position  (m)   & -0.1  &  0.7 & -5.2  & 2.5 \\
       \hline
       Downrange Velocity (m/s)  & 0.06 &  0.03   & -0.04   & 0.14 \\
       Crossrange Velocity (m/s)  & -0.01  &  0.04   &  -0.20 & 0.14 \\
       Elevation Velocity (m/s)  & -0.93 & 0.08   & -1.32   & -0.49 \\
       \hline
       Pitch (rad)  & -0.016 &  0.010   & -0.060   & 0.028 \\
       Roll  (rad)  &  -0.003 &  0.011   &  -0.043  & 0.050 \\
       \hline
       Rot. Velocity - Roll (rad/s) & -0.000 & 0.021    & -0.105   & 0.084 \\
       Rot. Velocity - Pitch (rad/s) & -0.005 &  0.013   &  -0.083  & 0.051 \\
       Rot. Velocity - Yaw (rad/s) & 0.000 & 0.000 & 0.000 & 0.000 \\
       \hline
       Glideslope & 22.02 & 17.77    & 7.50   & 851.54 \\
       \hline
       Fuel Consumed - 6-DOF RL Policy (kg) & 291  & 15    & 257   & 352 \\
       Fuel Consumed - 3-DOF RL Policy (kg) & 291  &  14   &  259 & 358 \\
       Fuel Consumed - 3-DOF DR/DV (kg) & 279  & 14    & 251   & 335 \\
   \end{tabular}
\end{table}

We compared the fuel efficiency of the 6-DOF RL agent to that of a 3-DOF RL agent using the same reward shaping function and to a 3-DOF DR/DV controller; the fuel statistics are at the bottom of Table~\ref{tab:TDS}.   We use the DR/DV results as a proxy for optimal performance, as the algorithm is energy-optimal for the case of unlimited thrust (although we limit the thrust for the comparison). Because DR/DV has an unacceptable terminal glideslope (less than 1), we used a piecewise trajectory with a single waypoint 15 m above the landing site, similar to the approach we used for the RL policy, which achieved a minimum glideslope of close to 8. This added about 20 kg to the DR/DV fuel consumption. We see a 4\% increase in fuel consumption for the 6-DOF RL agent as compared to 3-DOF DR/DV. To put this increase in perspective, note that tracking the DR/DV trajectory in a 6-DOF environment would certainly increase full consumption. Finally, note that the 6-DOF RL agent achieves fuel efficiency close to that of the 3-DOF RL agent; this tells us that the selected reward shaping function has a critical impact on fuel efficiency. There is certainly room for improvement here, as the exponential decrease in the magnitude of the target velocity is probably sub-optimal. Indeed, an optimal trajectory with an initial position far from the target landing site might accelerate towards the target to quickly close the downrange and crossrange distance in order to reduce trajectory time and use less fuel to maintain altitude.

Divert functionality was tested by running 5000 Monte Carlo simulations over the same initial conditions, but triggering a divert of 800 m downrange and 800 m crossrange  when the lander reached an altitude of 1500 m. On average, the divert maneuver resulted in a 30 kg average increase in fuel consumption but otherwise did not impact performance.

\begin{figure}[!h]
\begin{center}
\psfrag{Position (m)}[][]{\scriptsize{Position (m)}}
\psfrag{Altitude (m)}[][]{\scriptsize{Altitude (m)}}
\psfrag{Velocity (m/s)}[][]{\scriptsize{Velocity (m/s)}}
\psfrag{Attitude (Deg)}[][]{\scriptsize{Attitude (Deg)}}
\psfrag{Angular Velocity (Deg/s)}[][]{\scriptsize{Angular Velocity (Deg/s)}}
\psfrag{Thrust (N)}[][]{\scriptsize{Thrust (N)}}
\psfrag{Norm Downrange-Crossrange (m)}[][]{\scriptsize{Norm Downrange-Crossrange (m)}}
\psfrag{Time (s)}[][]{\scriptsize{Time (s)}}
\psfrag{roll}[][]{\scriptsize{roll}}
\psfrag{pith}[][]{\scriptsize{pitch}}
\psfrag{yaw}[][]{\scriptsize{yaw}}
\psfrag{x}[][]{\scriptsize{$x$}}
\psfrag{y}[][]{\scriptsize{$y$}}
\psfrag{z}[][]{\scriptsize{$z$}}
\psfrag{vx}[][]{\scriptsize{$v_x$}}
\psfrag{vy}[][]{\scriptsize{$v_y$}}
\psfrag{vz}[][]{\scriptsize{$v_z$}}
\psfrag{Tx}[][]{\scriptsize{$T_x$}}
\psfrag{Ty}[][]{\scriptsize{$T_y$}}
\psfrag{Tz}[][]{\scriptsize{$T_z$}}
\psfrag{wx}[][]{\scriptsize{$\omega_x$}}
\psfrag{wy}[][]{\scriptsize{$\omega_y$}}
\psfrag{wz}[][]{\scriptsize{$\omega_z$}}
\includegraphics[height=12cm]{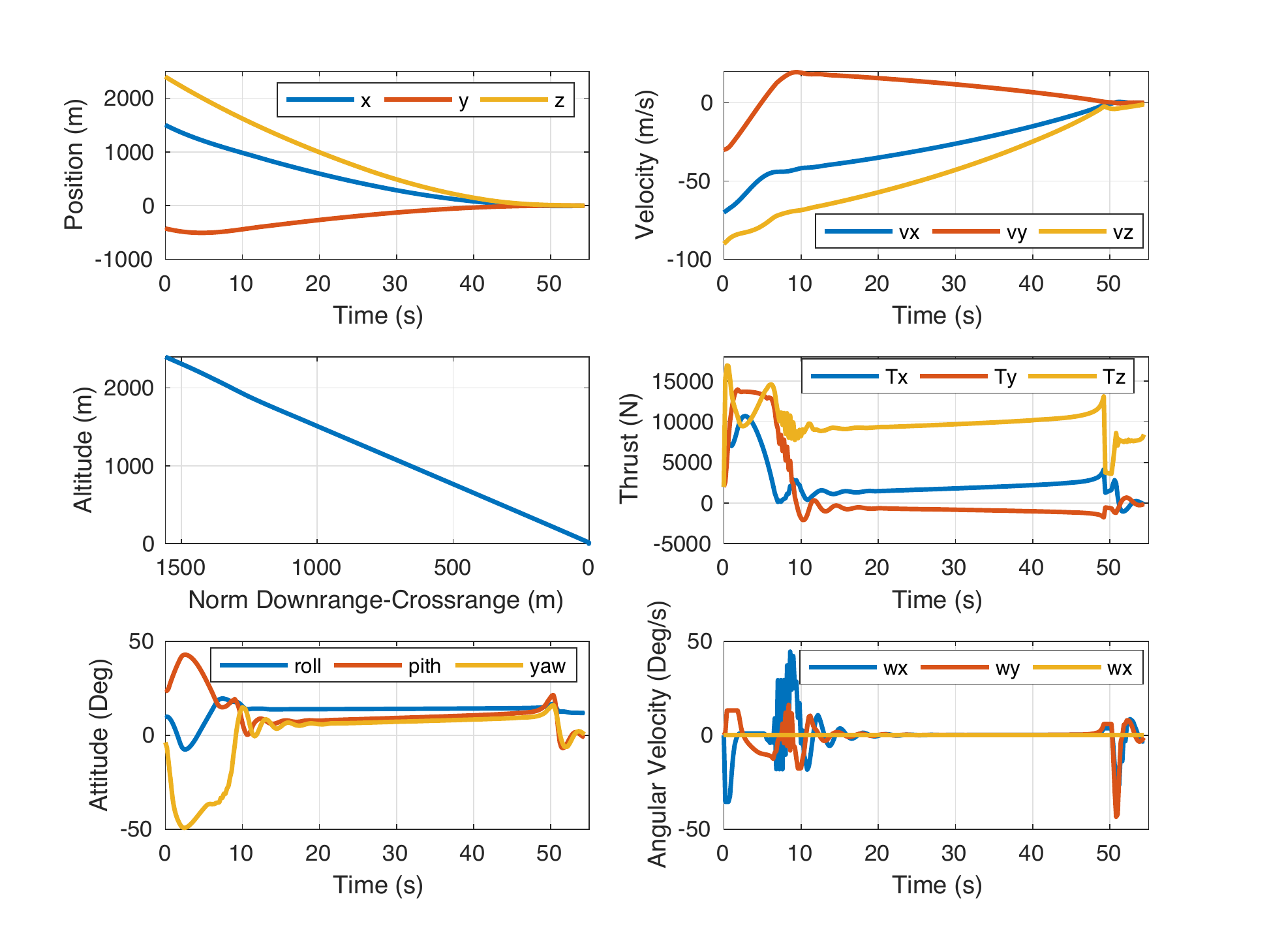}
\caption{6-DOF Sample Trajectory.}
\label{fig:st1}
\end{center}
\end{figure}

\begin{figure}[!h]
\begin{center}
\psfrag{Position (m)}[][]{\scriptsize{Position (m)}}
\psfrag{Altitude (m)}[][]{\scriptsize{Altitude (m)}}
\psfrag{Velocity (m/s)}[][]{\scriptsize{Velocity (m/s)}}
\psfrag{Attitude (Deg)}[][]{\scriptsize{Attitude (Deg)}}
\psfrag{Angular Velocity (Deg/s)}[][]{\scriptsize{Angular Velocity (Deg/s)}}
\psfrag{Thrust (N)}[][]{\scriptsize{Thrust (N)}}
\psfrag{Norm Downrange-Crossrange (m)}[][]{\scriptsize{Norm Downrange-Crossrange (m)}}
\psfrag{Time (s)}[][]{\scriptsize{Time (s)}}
\psfrag{x}[][]{\scriptsize{$x$}}
\psfrag{y}[][]{\scriptsize{$y$}}
\psfrag{z}[][]{\scriptsize{$z$}}
\psfrag{vx}[][]{\scriptsize{$v_x$}}
\psfrag{vy}[][]{\scriptsize{$v_y$}}
\psfrag{vz}[][]{\scriptsize{$v_z$}}
\psfrag{Tx}[][]{\scriptsize{$T_x$}}
\psfrag{Ty}[][]{\scriptsize{$T_y$}}
\psfrag{Tz}[][]{\scriptsize{$T_z$}}
\includegraphics[height=10cm]{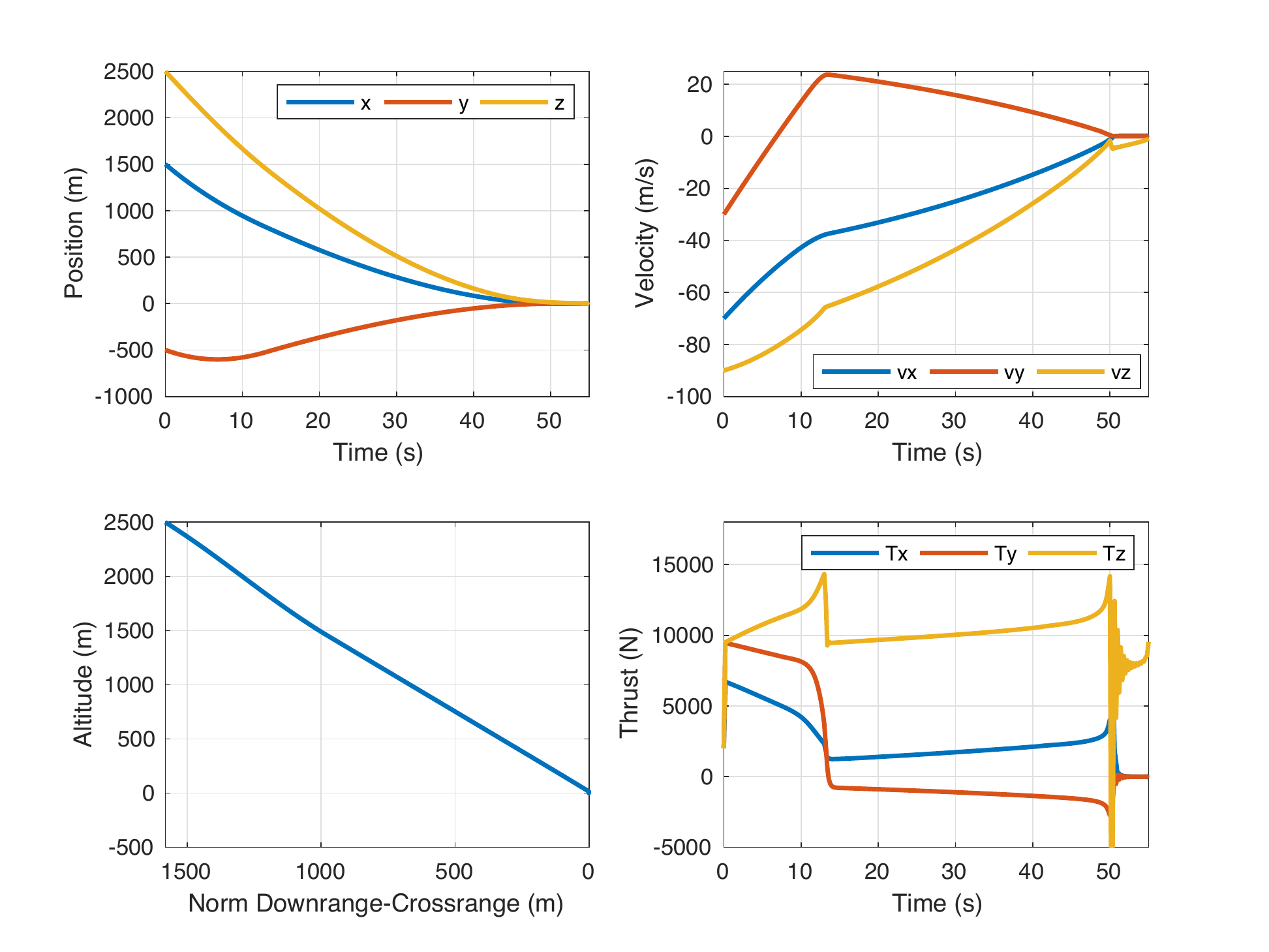}
\caption{Same Initial Conditions using 3-DOF policy.}
\label{fig:st2}
\end{center}
\end{figure}

A sample trajectory is plotted in Figure \ref{fig:st1}, where $x$, $y$, and $z$ are the downrange, crossrange, and altitude trajectory components in the target-centered reference frame.  Thrust is shown in the inertial frame. The left side plot that is second from the top gives the altitude as a function of the norm of the cross range and downrange position. This plot is from initial conditions of 1500 m to -70 m/s downrange, -500 m to -30 m/s crossrange, and 2400m to -90 m/s elevation. For comparison, we also plot (Figure \ref{fig:st2}) a trajectory from the 3-DOF RL policy that starts from the same initial conditions. Note that there are a couple of places where the system exhibits small oscillations in the lander's thrust vector. Training the agent longer possibly have a smoothing effect on the resulting solution.

From Figures \ref{fig:st1} and \ref{fig:st2},  it is apparent that the 3-DOF and 6-DOF position and velocity trajectories are almost identical. The thrust differs toward the start and end of the landing due to the need to adjust attitude in the 6-DOF case in order to change the thrust direction, although overall the thrust trajectories are similar. Specifically, since the simulations begin with an average pitch of 45 degrees, the policy rotates the lander to a smaller pitch angle in order to allow a burn with the thrust vector pointing downwards to reduce the magnitude of the vertical velocity component. In addition, the policy must rotate the lander to allow a burn that reduces the crossrange velocity until the lander is on a trajectory pointing towards the target. Towards the end of the trajectory, where the lander must transition to a vertical descent, there is another area where the thrust trajectories differ.

\begin{figure}[!h]
\begin{center}
\subfigure[3-DOF Trajectories]{
\psfrag{Elevation (m)}[][]{\scriptsize{Elevation (m)}}
\psfrag{Crossrange (m)}[][]{\scriptsize{Crossrange (m)}}
\psfrag{Downrange (m)}[][]{\scriptsize{Downrange (m)}}
\includegraphics[height=6cm]{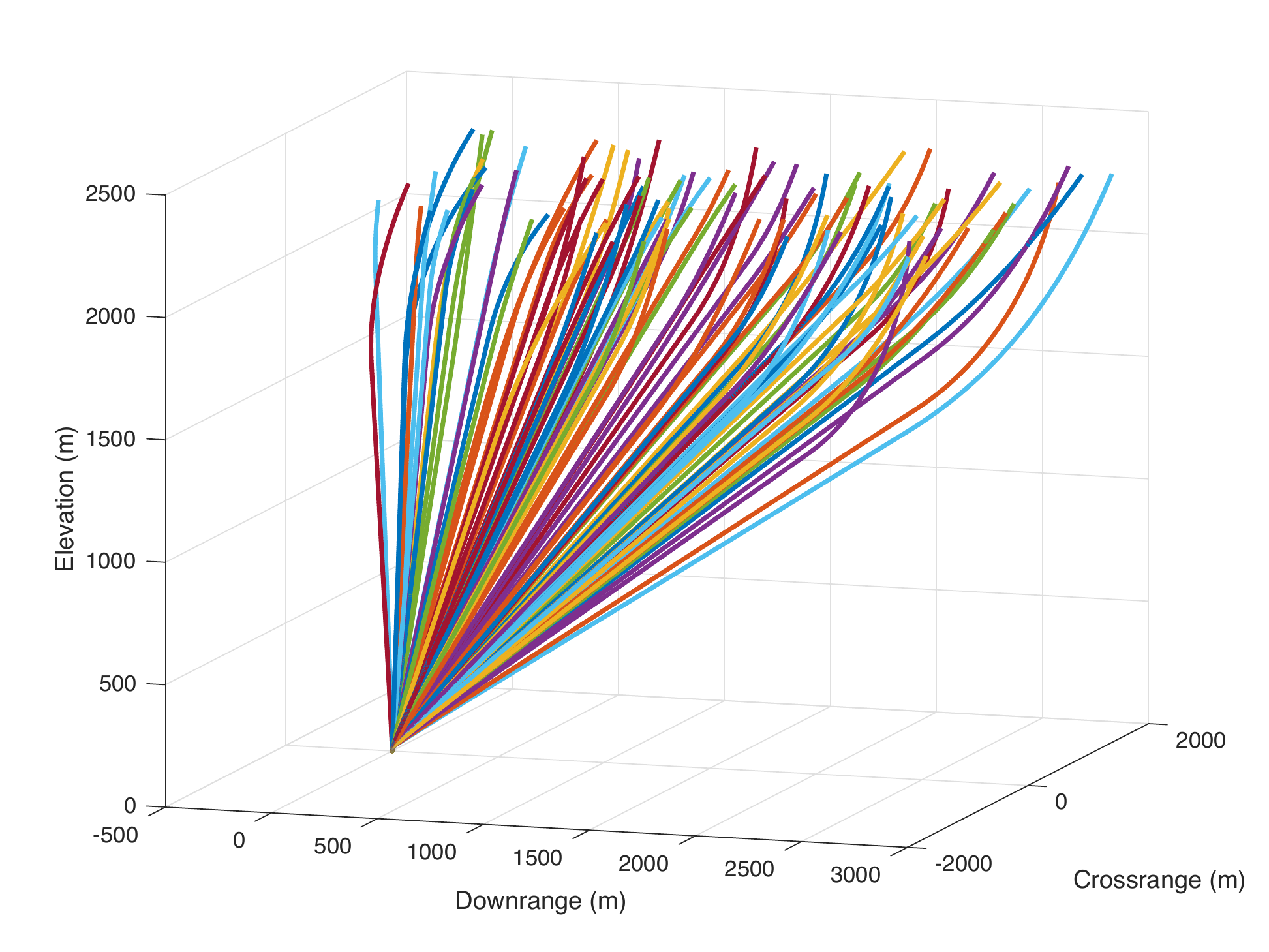}\label{fig:extT3dof}}
\subfigure[6-DOF Trajectories]{
\psfrag{Elevation (m)}[][]{\scriptsize{Elevation (m)}}
\psfrag{Crossrange (m)}[][]{\scriptsize{Crossrange (m)}}
\psfrag{Downrange (m)}[][]{\scriptsize{Downrange (m)}}
\includegraphics[height=6cm]{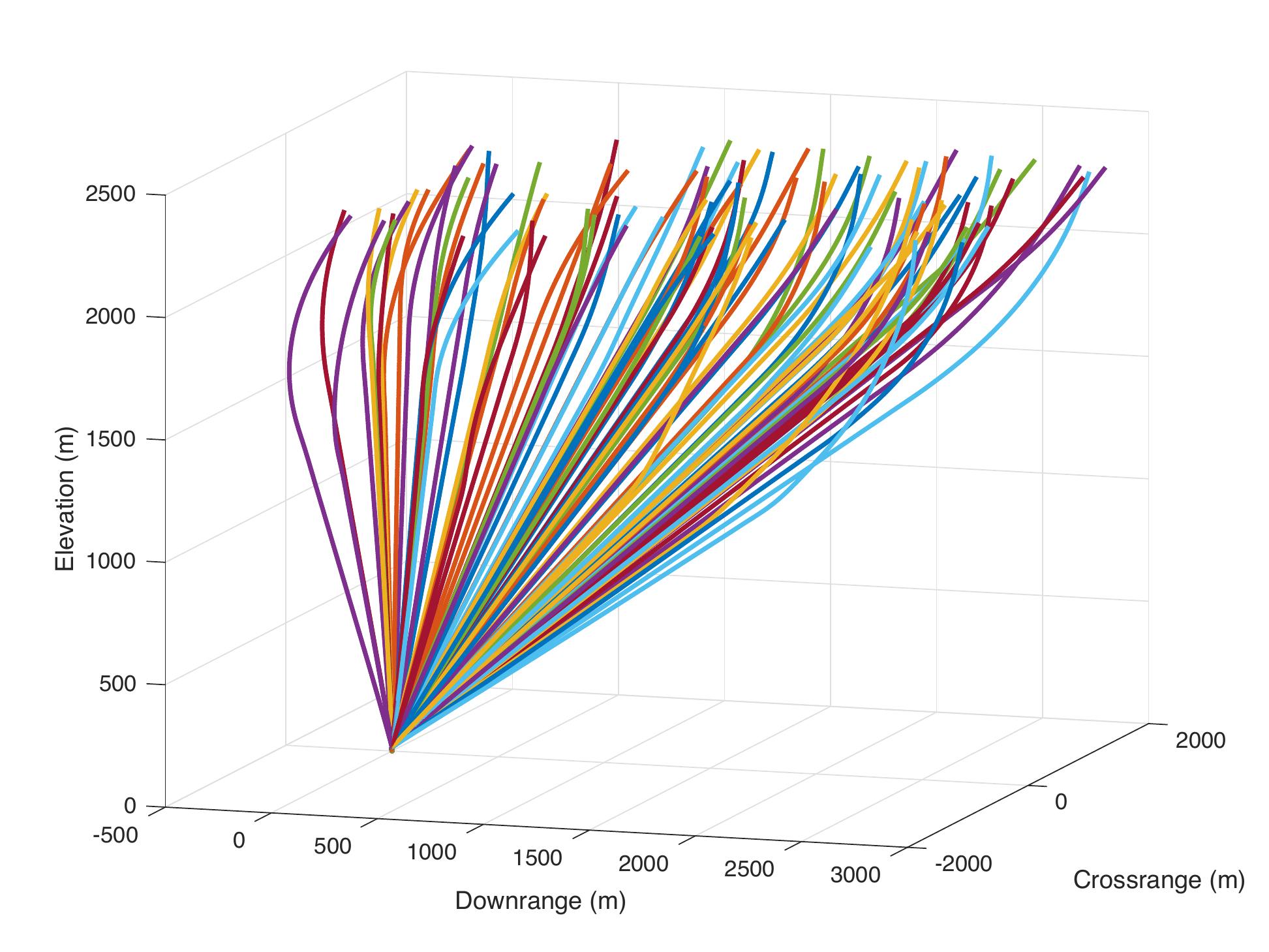}\label{fig:extT6dof}}
\caption{100 Sample Trajectories over 9 square km Deployment Ellipse.}
\label{fig:extT}
\end{center}
\end{figure}

We also re-tested the policy over two extended deployment ellipses, as shown in Table \ref{tab:EIC}, where we extend the deployment ellipse to 9 $\mathrm{km}^{2}$, and again to 12 $\mathrm{km}^{2}$. Here we use the same environmental noise and mass uncertainty as in the testing of the policy over the 4 $\mathrm{km}^{2}$ deployment ellipse. Note that in order to obtain satisfactory performance over the 12 $\mathrm{km}^{2}$ deployment ellipse, we had to raise the altitude of the deployment ellipse to 3000m. Importantly, this is the same policy that was optimized over the initial conditions given in Table \ref{tab:IC}, i.e., a 4 $\mathrm{km}^{2}$ deployment ellipse. Landing statistics over 20,000 episodes were similar to that given in Table \ref{tab:TDS}, but average and maximum fuel consumption increased. Since the reward shaping function does not require the lander's full translational state, the policy generalizes quite well to regions of state space not experienced during optimization. Figures \ref{fig:extT3dof} and \ref{fig:extT6dof} illustrate 100 randomly sampled trajectories using the 9 $\mathrm{km}^{2}$ deployment ellipse for the 3-DOF and 6-DOF policies, respectively. 

\label{fig:compare}

\begin{table}[!h]
	\fontsize{10}{10}\selectfont
    \caption{Lander Extended Initial Conditions for Optimization.}
   \label{tab:EIC}
        \centering 
   \newcolumntype{R}{>{\raggedleft\arraybackslash}p{1.8cm}}
   \begin{tabular}{l | R | R | R | R } 
      \hline 
       & \multicolumn{2}{c}{Velocity}\vline & \multicolumn{2}{c}{Position}\\
       \hline
       & min (m/s) & max (m/s) & min (m) & max (m) \\
       \hline
       \multicolumn{5}{c}{9 km Deployment Ellipse}\\
      \hline
      Downrange      & -70 & -10 & 0 & 3000\\
      Crossrange       & -30  & 30 & -1500 & 1500 \\
      Elevation     & -90 & -70 & 2400 & 2500 \\
      \hline
      \multicolumn{5}{c}{12 km Deployment Ellipse}\\
      \hline
      Downrange      & -70 & -10 & 0 & 4000\\
      Crossrange       & -30  & 30 & -1500 & 1500 \\
      Elevation     & -90 & -70 & 2900 & 3100 \\
   \end{tabular}
\end{table}

\begin{table}[!h]
	\fontsize{10}{10}\selectfont
    \caption{Extended Range Fuel Consumption.}
   \label{tab:ext_fuel}
   \centering 
   \newcolumntype{R}{>{\raggedleft\arraybackslash}p{1.5cm}}
\begin{tabular}{l | R | R | R | R } 
   
      \hline
        & Mean  & Std. Dev.  &  Min & Max \\
      \hline
       \multicolumn{5}{c}{9 km Deployment Ellipse}\\
      \hline
       Fuel Consumed - 6-DOF RL Policy (kg) &  308 & 25    & 257   & 412 \\
       Fuel Consumed - 3-DOF RL (kg) & 309 &  25   &  260  &  400\\
       Fuel Consumed - 3-DOF DR/DV (kg) & 297 & 23 & 252 & 381 \\
       \hline
       \multicolumn{5}{c}{12 km Deployment Ellipse}\\
      \hline
      Fuel Consumed - 6-DOF RL Policy (kg) &  340 & 31    & 282   & 468 \\
       Fuel Consumed - 3-DOF RL (kg) & 341 &  30   &  285  &  437\\
       Fuel Consumed - 3-DOF DR/DV (kg) & 319 & 27 & 268 & 414
   \end{tabular}
\end{table}


\begin{figure}[!h]
\begin{center}
\subfigure[3-DOF Trajectories]{
\psfrag{3 DoF}[][]{\scriptsize{3-DOF}}
\psfrag{6 DoF}[][]{\scriptsize{6-DOF}}
\psfrag{GPOPS}[][]{\scriptsize{GPOPS}}
\psfrag{Mass (kg)}[][]{\scriptsize{Mass (kg)}}
\psfrag{Time (s)}[][]{\scriptsize{Time (s)}}
\includegraphics[height=6cm]{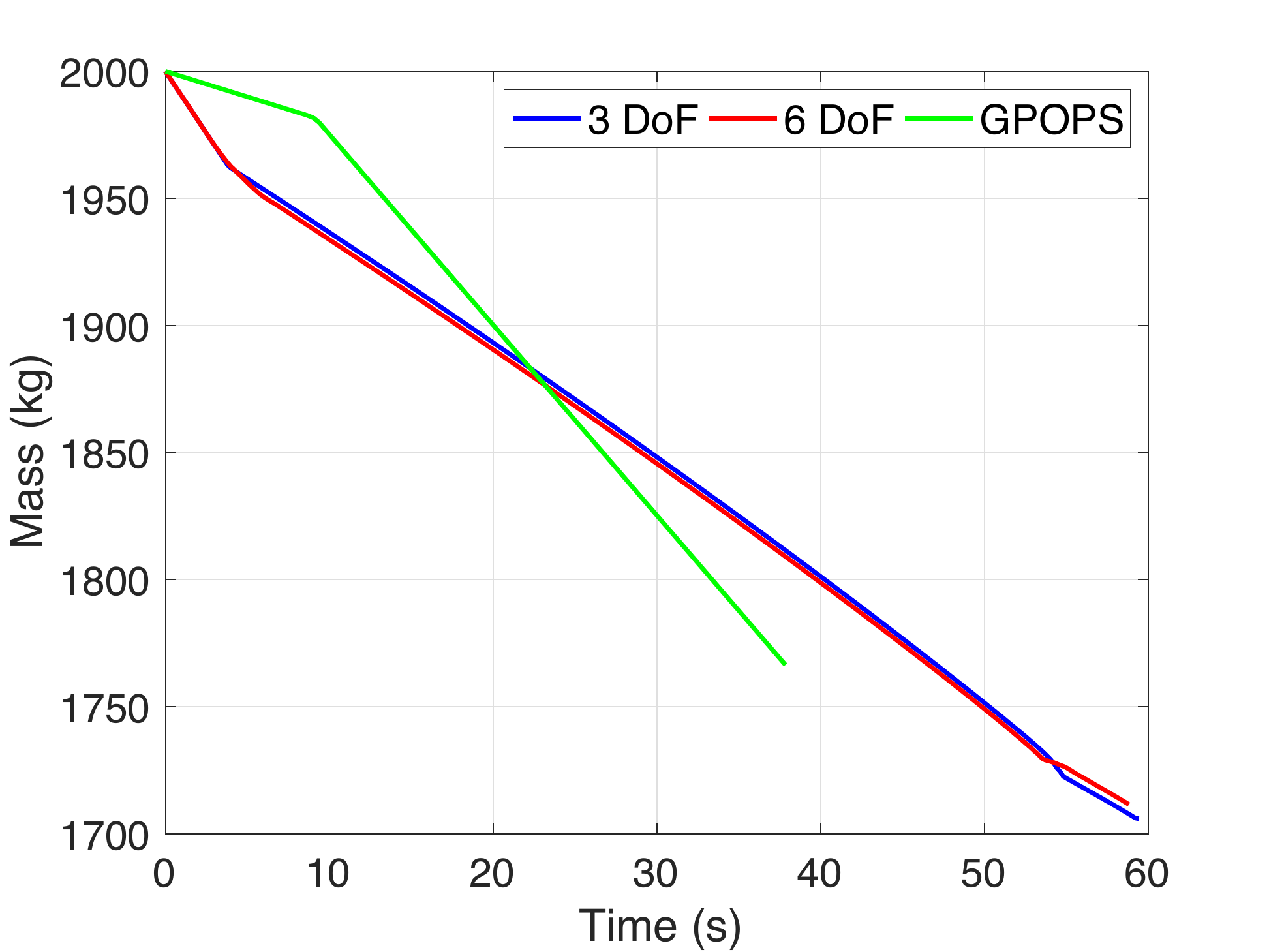}}
\subfigure[6-DOF Trajectories]{
\psfrag{3 DoF}[][]{\scriptsize{3-DOF}}
\psfrag{6 DoF}[][]{\scriptsize{6-DOF}}
\psfrag{GPOPS}[][]{\scriptsize{GPOPS}}
\psfrag{Elevation (m)}[][]{\scriptsize{Elevation (m)}}
\psfrag{Crossrange (m)}[][]{\scriptsize{Crossrange (m)}}
\psfrag{Downrange (m)}[][]{\scriptsize{Downrange (m)}}
\includegraphics[height=6cm]{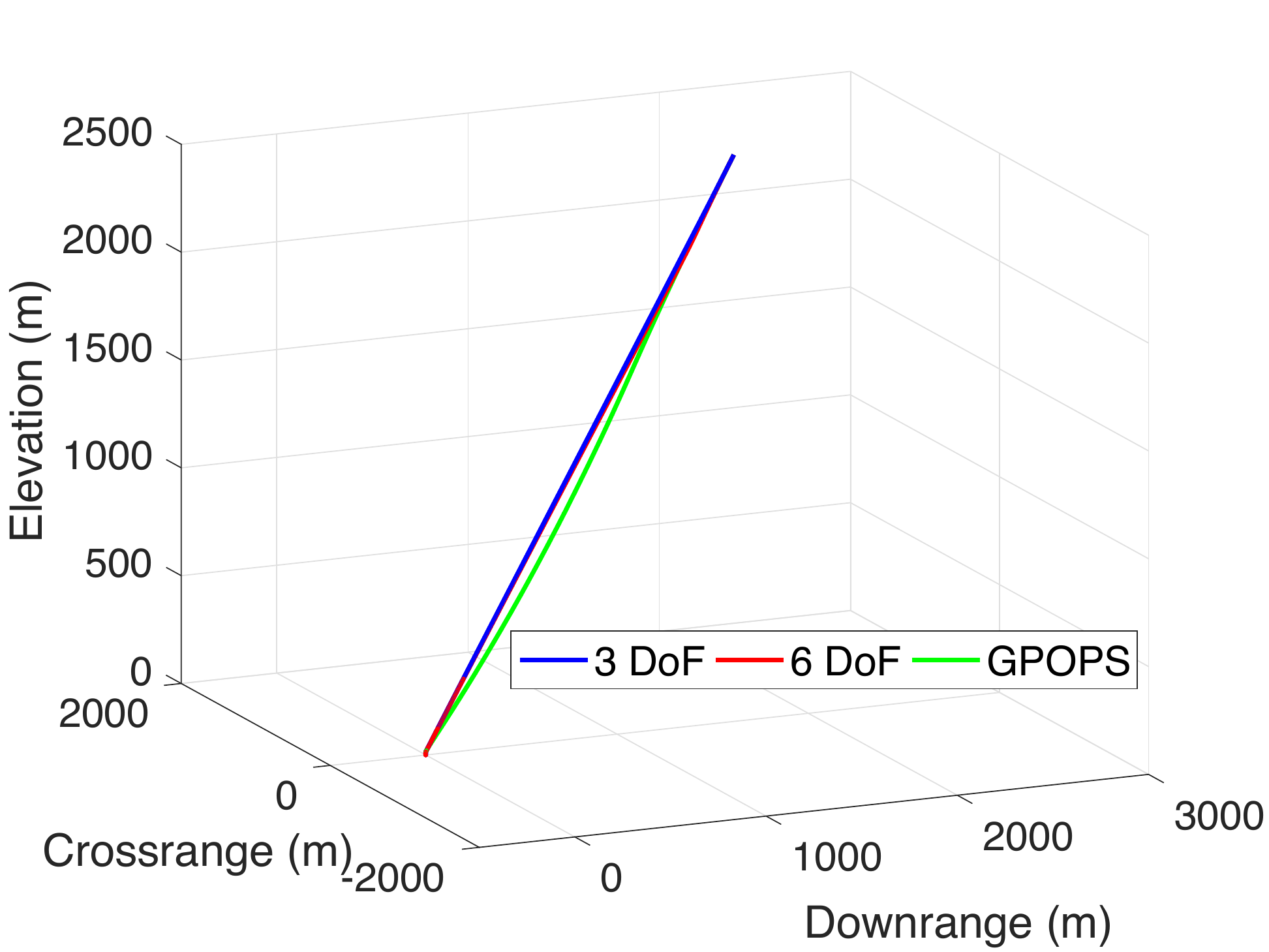}}
\caption{Comparison of 3-DOF and 6-DOF policy with GPOPS solution.}
\label{fig:compare}
\end{center}
\end{figure}

\section{Comparison with GPOPS Solution}

\begin{figure*}[!ht]
\begin{centering}
\subfigure[3-DOF Fuel-Mass Performance]{
\psfrag{Divert range (m)}[][]{\scriptsize{Divert range (m)}}
\psfrag{Velocity (m/s)}[][]{\scriptsize{Velocity (m/s)}}
\psfrag{Delta Mass (kg)}[][]{\scriptsize{Final $\Delta m$ (kg)}}
\psfrag{mass=0.026r-0.961v+340}[][]{\scriptsize{$\Delta m$=0.026$\|{\bf r}\|$-0.961$\|{\bf v}\|$+340}}
\includegraphics[keepaspectratio, width=0.45\textwidth]{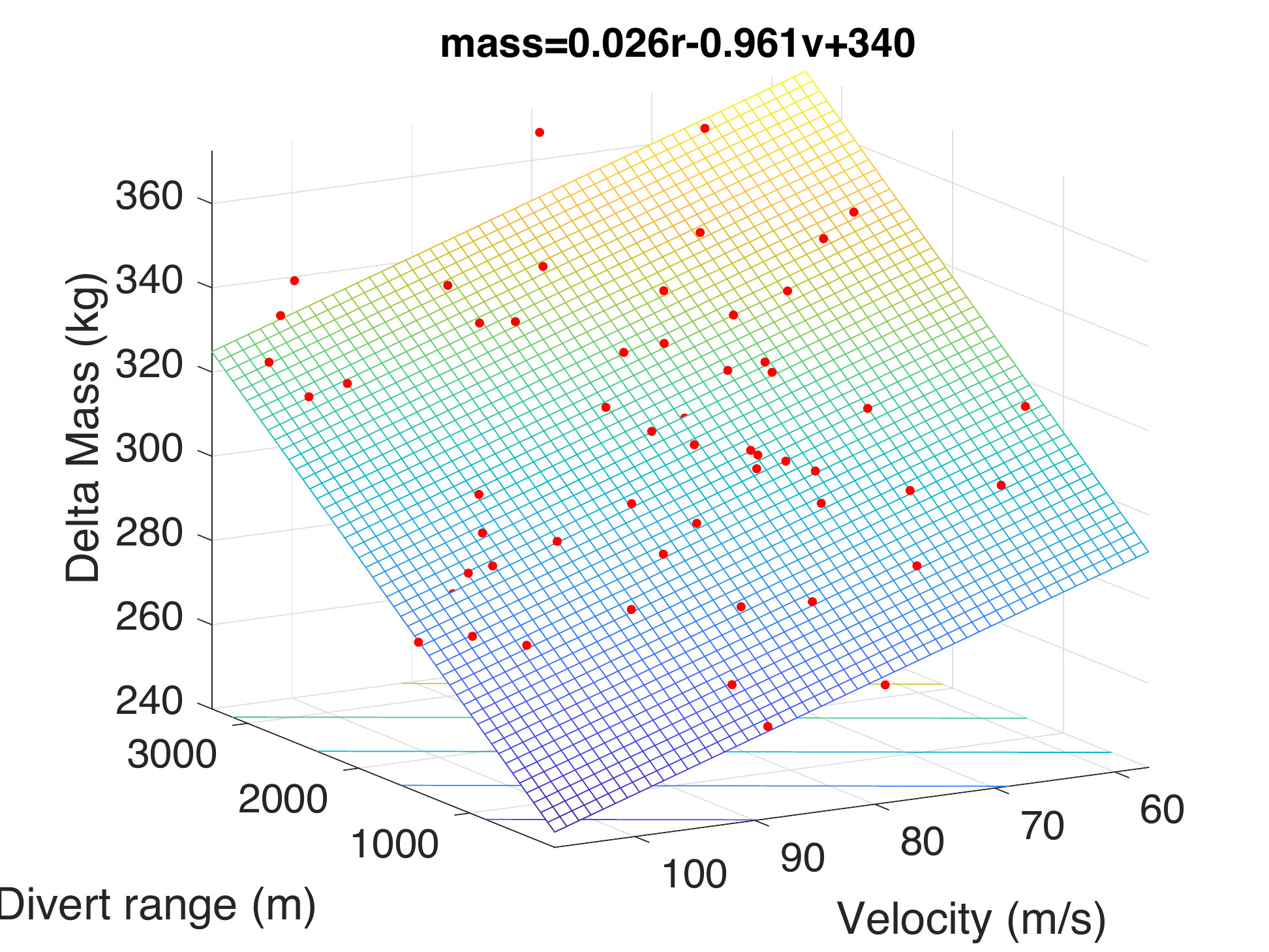}\label{fig:3dof_fuel}}
\subfigure[6-DOF Fuel-Mass Performance]{
\psfrag{Divert range (m)}[][]{\scriptsize{Divert range (m)}}
\psfrag{Velocity (m/s)}[][]{\scriptsize{Velocity (m/s)}}
\psfrag{Delta Mass (kg)}[][]{\scriptsize{Final $\Delta m$ (kg)}}
\psfrag{mass=0.025r-1.091v+351}[][]{\scriptsize{$\Delta m$=0.025$\|{\bf r}\|$-1.091$\|{\bf v}\|$+351}}
\includegraphics[keepaspectratio, width=0.45\textwidth]{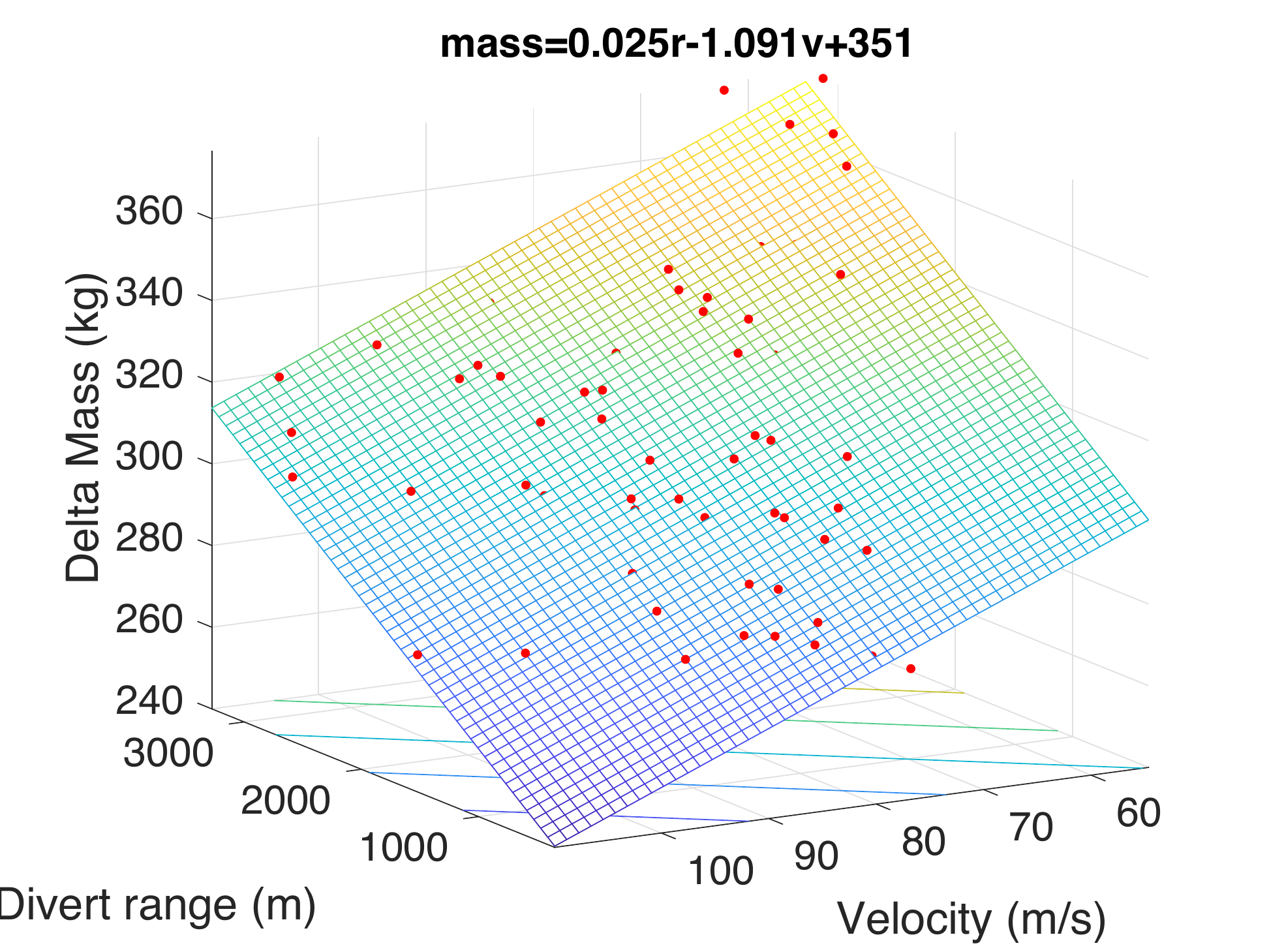}\label{fig:6dof_fuel}}
\\
\subfigure[3-DOF Fuel-Mass Along the Trajectories]{
\psfrag{Divert range (m)}[][]{\scriptsize{Divert range (m)}}
\psfrag{Velocity (m/s)}[][]{\scriptsize{Velocity (m/s)}}
\psfrag{Delta Mass (kg)}[][]{\scriptsize{$\Delta m$ (kg)}}
\psfrag{mass=0.026r-0.961v+340}[][]{\scriptsize{$\Delta m$=0.026$\|{\bf r}\|$-0.961$\|{\bf v}\|$+340}}
\includegraphics[keepaspectratio, width=0.45\textwidth]{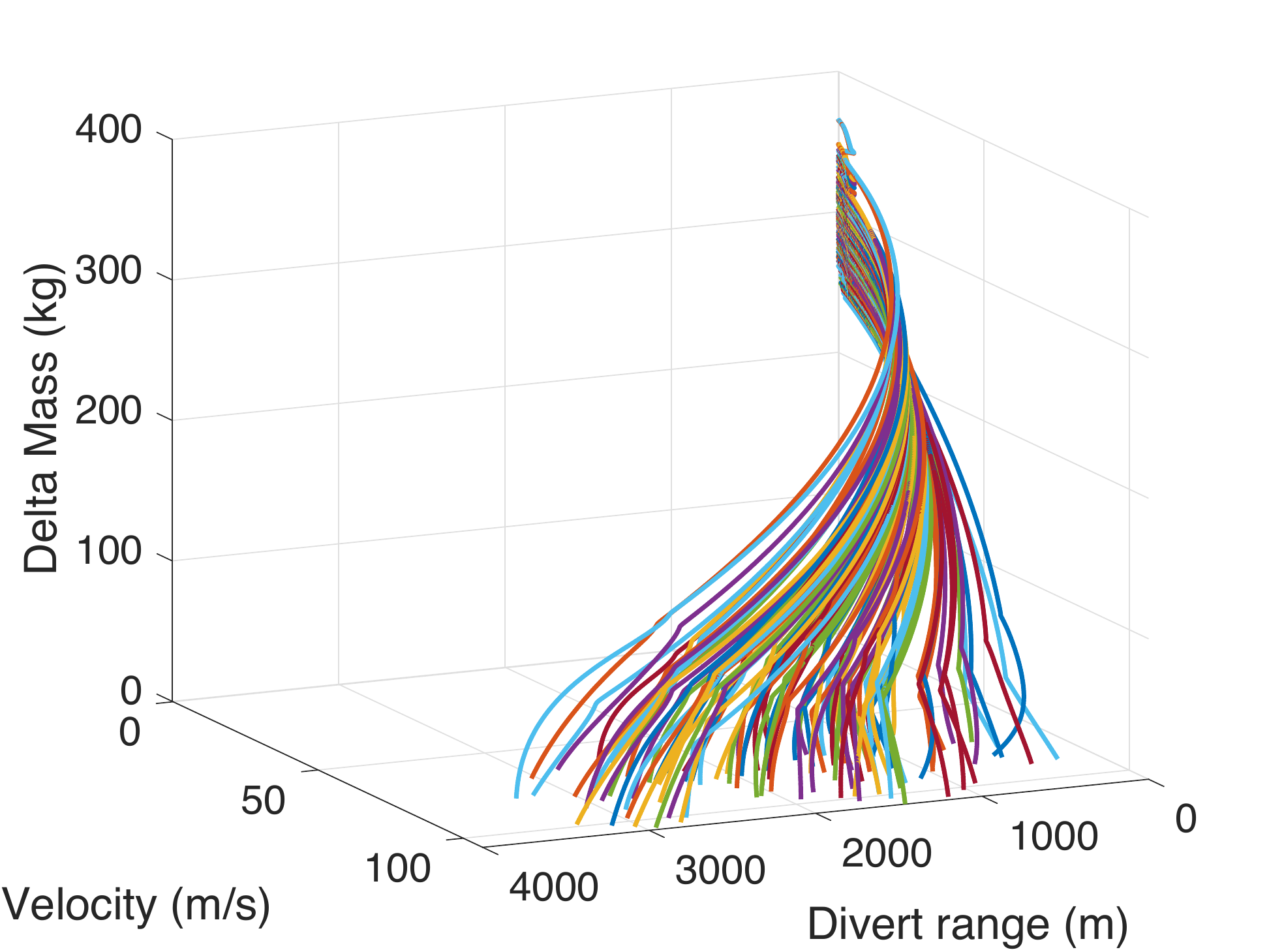}\label{fig:mass_along_trajectories_3dof}}
\subfigure[6-DOF Fuel-Mass Along the Trajectories]{
\psfrag{Divert range (m)}[][]{\scriptsize{Divert range (m)}}
\psfrag{Velocity (m/s)}[][]{\scriptsize{Velocity (m/s)}}
\psfrag{Delta Mass (kg)}[][]{\scriptsize{$\Delta m$ (kg)}}
\psfrag{mass=0.025r-1.091v+351}[][]{\scriptsize{$\Delta m$=0.025$\|{\bf r}\|$-1.091$\|{\bf v}\|$+351}}
\includegraphics[keepaspectratio, width=0.45\textwidth]{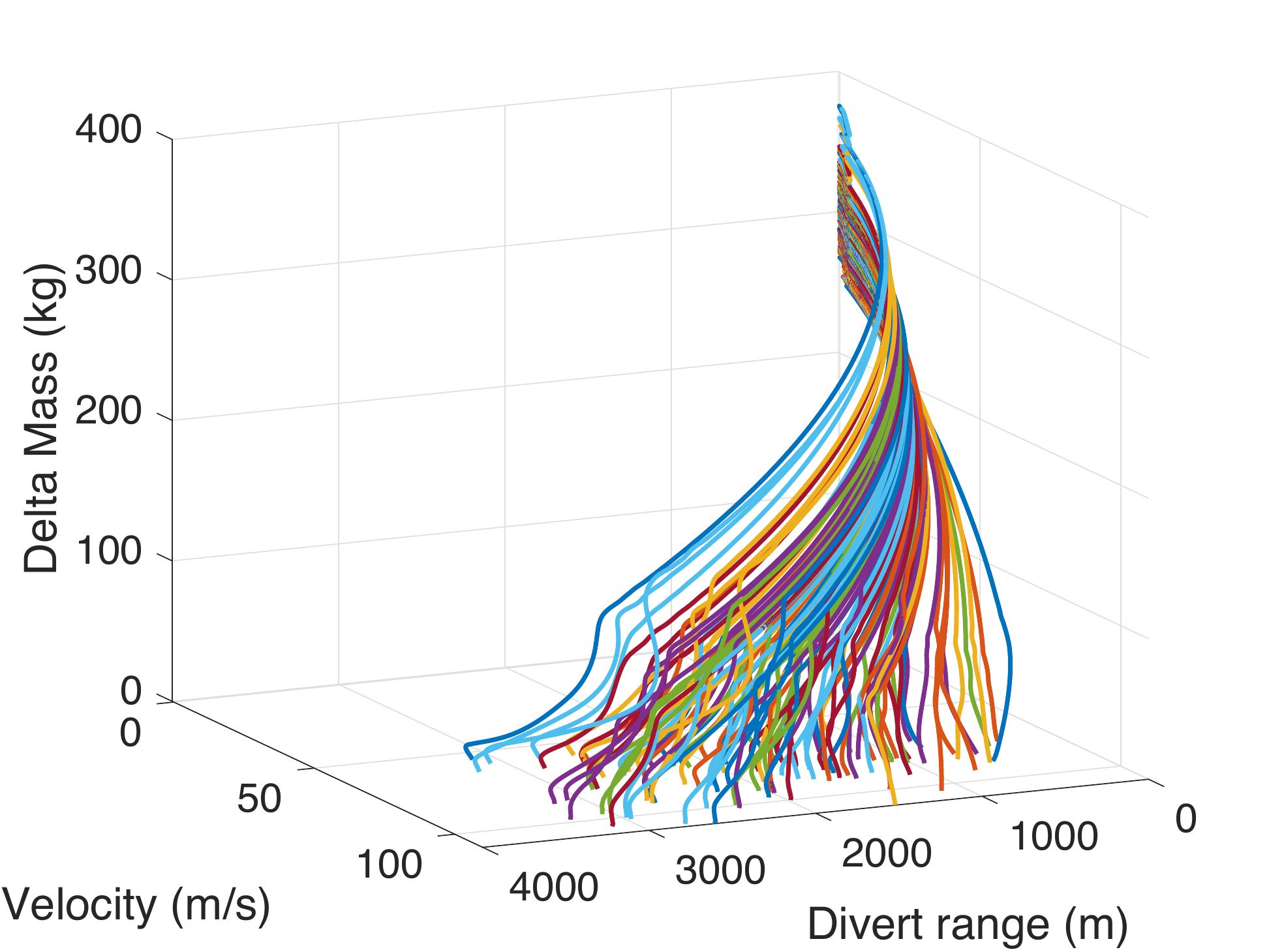}\label{fig:mass_along_trajectories_6dof}}
\caption{6-DOF and 3-DOF Fuel-Mass Performance.}\label{fuel}
\end{centering}
\end{figure*}

Figure \ref{fig:compare} shows experimental results comparing the 3-DOF and 6-DOF agents to solution provided by GPOPS [\citenum{rao2010algorithm}], an  optimal control solver, using a 3-DOF problem formulation with the same initial conditions used to generate Figures \ref{fig:st1} and \ref{fig:st2}. The GPOPS solution, which also required a vertical descent over the final 15 m of altitude loss, had a fuel consumption of 250 kg, as compared to 290 kg for the 6-DOF and 3-DOF policies, making the RL policy fuel consumption 18 percent higher than GPOPS optimal. First, note that the GPOPS trajectories are open loop, and when combined with a trajectory tracking controller the fuel consumption will be higher. Second, since the 6-DOF and 3-DOF policies have almost identical fuel consumption, we can attribute the difference in fuel efficiency between the RL policies and optimal to the reward shaping function used during optimization, which although effective, is not optimal. Consequently, future work to improve the fuel efficiency of the RL derived integrated guidance and control system should focus on the reward shaping function. Since this function takes the form of a velocity field, it may prove productive to learn a reward shaping function parameterized as a neural network from optimal trajectories. 
Figures \ref{fig:3dof_fuel}, \ref{fig:6dof_fuel}, \ref{fig:mass_along_trajectories_3dof}, and \ref{fig:mass_along_trajectories_6dof} show the fuel performance for the 3-DOF and 6-DOF policies over the 9 square kilometer deployment ellipse. Interestingly, the zero divert fuel consumption in Figures \ref{fig:3dof_fuel} and \ref{fig:6dof_fuel} is around 240kg, close to that of the MSL if we subtract fuel used in the sky-crane maneuver.

\section{Conclusion}

We have applied a novel approach to the Mars landing powered descent phase problem using an integrated guidance and control system developed using reinforcement learning, and validated the approach through Monte Carlo simulation.  Coupled with a navigation system such as that described in ~[\citenum{gaudet2014navigation}], this guidance and control system can achieve pinpoint accuracy and a soft landing with minimal deviation from an ideal landing attitude and rotational velocity, with large divert distance capability. As compared to current practice, this system provides a significant increase in landing precision.  The policy was shown to be robust to noise and parameter uncertainty. Compared to proposed systems such as that described in [\citenum{accikmecse2013lossless}], this system has the advantage of not requiring a cone-shaped glideslope constraint, allowing the targeting of locations such as the bottom of a deep crater. The computational requirements of running the policy are modest, with only four matrix multiplications required to map the estimated state from the navigation system to a body frame thrust command, and then a trained policy should have no problem running on the current generation of flight computers.

\section{Acknowledgment}

 We  borrowed several techniques from the PPO2 implementation at open-AI gym baselines. Our Python code adapted functions from ``Analytical Mechanics of Space Systems" by Schaub and Junkins [\citenum{junkins2009analytical}].

\bibliographystyle{AAS_publication}   
\bibliography{references}   

\end{document}